\documentclass[10pt]{article}
\usepackage{amsmath}
\usepackage{times}
\usepackage{graphicx}
\usepackage[authoryear,sort]{natbib}

\newcommand{\captionfonts}{\normalsize}

\makeatletter  
\long\def\@makecaption#1#2{%
  \vskip\abovecaptionskip
  \sbox\@tempboxa{{\captionfonts #1: #2}}%
  \ifdim \wd\@tempboxa >\hsize
    {\captionfonts #1: #2\par}
  \else
    \hbox to\hsize{\hfil\box\@tempboxa\hfil}%
  \fi
  \vskip\belowcaptionskip}
\makeatother   

\DeclareMathAlphabet{\mathitbf}{OT1}{cmr}{bx}{it} 
\newcommand{\mat}[1]{\mathitbf{#1}} 
\newcommand{\oned}{\mbox{\textsc{1-d}}} 
\newcommand{\twod}{\mbox{\textsc{2-d}}}  
\newcommand{\tp}{{\scriptscriptstyle\top}} 
\newcommand{\eqand}{\quad\text{and}\quad} 
\newcommand{\eqwhere}{\quad\text{where}\quad} 

\begin{document}
\hspace{13.9cm}1

\ \vspace{20mm}\\

{\LARGE\bf A Differential Model of the Complex Cell}

\ \\
{\bf \large Miles Hansard$^{\displaystyle 1}$ and Radu Horaud$^{\displaystyle 1}$}\\
{$^{\displaystyle 1}$\textsc{Inria} Rh\^{o}ne-Alpes, 655 Avenue de l'Europe, Montbonnot, France 38330.}\\



\thispagestyle{empty}

\begin{center} {\bf Abstract} \end{center}
The receptive fields of simple cells in the visual cortex can be 
understood as linear filters. These filters can be modelled by 
Gabor functions, or by Gaussian derivatives. Gabor functions can also be 
combined in an `energy model' of the complex cell 
response. This paper proposes an alternative model of the complex cell, based 
on Gaussian derivatives. It is most important to account for the insensitivity 
of the complex response to small shifts of the image. The new model uses a 
linear combination of the first few derivative filters, at a single position, 
to approximate the first derivative filter, at a series of adjacent positions.
The maximum response, over all positions, gives a signal that is insensitive 
to small shifts of the image. This model, unlike previous approaches, is
based on the scale space theory of visual processing. In particular, 
the complex cell is built from filters that respond to the \twod\ differential
structure of the image. The computational aspects of the new model are studied 
in one and two dimensions, using the steerability of the Gaussian derivatives.
The response of the model to basic images, such as edges and gratings,
is derived formally. The response to natural images is also evaluated,
using statistical measures of shift insensitivity. The relevance of
the new model to the cortical image representation is discussed.

\section{Introduction}
\label{sec:intro}

\noindent
It is useful to distinguish between simple and complex cells in the visual 
cortex, as proposed by \cite{hubel-1962}. The original classification 
is based on four characteristic properties of simple cells, as follows. 
Firstly, the receptive field has distinct excitatory and 
inhibitory subregions. Secondly, there is spatial summation within
each subregion. Thirdly, there is antagonism between the excitatory and 
inhibitory subregions. Fourthly, the response to any stimulus can be predicted 
from the receptive field map. Cells that do not show these characteristics can
be classified, by convention, as complex.

The distinction between simple and complex cells has endured, subject to certain
qualifications (although an alternative view is described by \citealp{mechler-2002a}).
In particular, it has been 
argued that the principle characteristic of complex cells is the \emph{phase 
invariance} of the response \citep{movshon-1978b,carandini-2006}. This means
that a complex cell, which is tuned to a particular orientation and spatial 
frequency, is not sensitive to the precise \emph{location} of the stimulus
within the receptive field \citep{kjaer-1997,mechler-2002b}. If the stimulus 
consists of a moving (or flickering)
grating, then phase invariance can be quantified by the relative modulation
$a_1/a_0$, where $a_1$ is the amplitude associated with the fundamental frequency 
of the response, and $a_0$ is the mean response 
\citep{movshon-1978b,devalois-1982,skottun-1991}. If this ratio is less than one, 
then the cell can be classified as complex.

The standard model of the simple cell is based on a linear filter that is localized
in position, spatial frequency and orientation. The output of this filter is subject
to a nonlinearity, such as squaring, and may also be normalized by the responses
of nearby simple cells. The theoretical framework of this model is well advanced 
(as reviewed by \citealp{dayan-2001,carandini-2005}), owing to the assumed linearity 
of the underlying spatial filters.
Although the physiology of the complex cell is increasingly well-understood,
(as reviewed by \citealp{spitzer-1988,alonso-1998,martinez-2003}), the appropriate theoretical 
framework is less clear. It is useful to adopt the distinction between `position'
and `phase' models that was made by \cite{fleet-1996} in the
analysis of binocular processing. It is \emph{invariance} to position or phase 
that is of interest in the present context.

A position-invariant model of the complex cell can be constructed from 
a set of simple cells, as described by \cite{hubel-1962}. Suppose, for
example, that the simple cells are represented by linear filters of a common 
orientation, but different spatial positions. If the responses of these filters 
are summed, then the corresponding complex cell will be tuned to an oriented element that
appears anywhere in the union of the simple cell receptive fields \citep{spitzer-1988}.
This scheme is the basis of the \emph{subunit model}, which was further developed
and tested by \cite{movshon-1978a,movshon-1978b}. The results
of these experiments are consistent with the idea that the complex response is based 
on a group of spatially \emph{linear} subunits. The most straightforward way to combine
the individual responses would be by rectifying and summing them, but the maximum could
also be taken \citep{riesenhuber-1999}. 
The main disadvantage of the subunit model is that it is \emph{too general} in its 
basic form. For example, it allows arbitrarily complicated receptive fields to be 
constructed, as there is no intrinsic constraint on the positions, orientations or 
spatial frequencies of the subunits.

Alternatively, a phase-invariant model can be constructed from a pair of filters 
of different shapes (\citep{adelson-1985}). If the filters 
have a quadrature relationship, then their responses to a sinusoidal stimulus
will differ in phase by $\pi/2$.
It follows that the sum of the squared outputs will
be invariant to the phase of the input. 
The application of this \emph{energy model} to spatial vision 
\citep{morrone-1988,emerson-1992,atherton-2002,wundrich-2004} is motivated 
by the observed phase-differences in the receptive fields of 
adjacent simple cells \cite{pollen-1981}. Indeed, these receptive fields can be
well-represented by odd and even Gabor functions 
\citep{daugman-1985,jones-1987,pollen-1983}.
There are two problems with the energy model, in the present context. Firstly, 
there is no generally agreed way to combine energy mechanisms across different 
frequencies and orientations (one approach is described by \citealp{fleet-1996}). 
This is an obstacle to the construction of mechanisms that show more complicated invariances,
such as those found in areas MT and MST \citep{orban-2008}. Secondly, the 
quadrature filters that are best-suited to the energy model are not convenient 
for the general description of \twod\ image-structure. The concept of phase 
itself becomes somewhat complicated in two or more dimensions \citep{felsberg-2001}, 
and the quadrature representation of more complex image-features, 
such as edge curvature, is unclear. It must be emphasized that \twod\ images
contain important structures (e.g.\ luminance saddle-points) that have no 
analogue in the \oned\ signals to which the energy model is ideally suited.
A realistic framework for spatial vision must be capable of representing the
full variety of \twod\ structures \citep{dobbins-1987,petitot-2003,shahar-2004}.

The present work is motivated by the difficulty of extending the energy model 
to more complicated \twod\ image-features and spatial transformations. These 
extensions require a representation of the local image \emph{geometry}, rather 
than the local phase and frequency structure.
The new approach, like the energy model, is based on a set of odd and even
linear filters that are located at the same position. The outputs of 
these filters are nonlinearly combined, again as in the energy model. 
The combination, however, involves the implicit construction of 
spatially-offset subunits. The complete model, in this sense, can be seen
as a re-formulation of the \cite{hubel-1962} scheme.

\subsection{Formal Overview}
\label{sec:present}

\noindent

A minimal overview of the new model will now be given. Let $S(\mat{x})$
be the original scalar image, where $\mat{x} = (x,y)^\tp$, and 
consider a spatial array of simple cells,
parameterized by preferred frequency and orientation. 
These receptive fields will be modelled
by $k$-th order directional derivatives of the Gaussian kernel,
\mbox{$G_k(\mat{x},\sigma,\theta) = (\mat{v}\cdot\nabla)^k G_0(\mat{x},\sigma)$}
where $\sigma$ is the spatial scale, $\theta$ is the orientation, and
$\mat{v} = (\cos\theta,\sin\theta)^\tp$. The simple cell representation 
$S_k(\mat{x},\sigma,\theta)$
is given by the convolution of these filters with the image:
\begin{equation}
S_k(\mat{x},\sigma,\theta) = G_k(\mat{x},\sigma,\theta)\star S(\mat{x}).
\label{eqn:simple}
\end{equation}
In particular, consider the magnitude of the first derivative signal, 
$|S_1(\mat{x},\sigma,\theta)|$. This will be large if there is a step-like
edge at $\mat{x}$, with the luminance boundary perpendicular to $\mat{v}$.
Now suppose that the edge is shifted by some amount in direction $\mat{v}$. 
This means that the magnitude $|S_1(\mat{x},\sigma,\theta)|$ will fall, but the nonlinear function 
\begin{equation}
C(\mat{x},\sigma,\theta) =
\underset{t}{\max}\mspace{4mu}
\bigl|
S_1(\mat{x}+t\mat{v},\sigma,\theta)
\bigr|, 
\eqwhere |t| \le \rho
\label{eqn:complex}
\end{equation}
will remain large, unless the shift exceeds the range $\rho$.
Equations (\ref{eqn:simple} \& \ref{eqn:complex}) will be the 
basic models of simple cells $S_k(\mat{x},\sigma,\theta)$ and
complex cells $C(\mat{x},\sigma,\theta)$ in this paper (full 
derivations are given in sec.~\ref{sec:offset}). 
The complex cell,
which inherits the scale and orientation tuning $(\sigma,\theta)$, has a
receptive field of radius $\rho$, centred on position $\mat{x}$. It can be seen that 
(\ref{eqn:complex}) is just a special case of the \cite{hubel-1962} subunit
model, with simple cells distributed along the spatial 
axis $\mat{v}$, and `max' being the combination rule. It has already been
argued, in section \ref{sec:intro}, that this model is too general. 
For example, there is no natural limit on the size $\rho$ of the complex 
receptive field in (\ref{eqn:complex}).

Suppose, however, that access to the 
first-order directional structure
\emph{around} position $\mat{x}$ is replaced by
access to the higher-order directional structure \emph{at}
position $\mat{x}$.
Mathematically, this means that the function 
$S_1(\mat{x}+t\mat{v},\sigma,\theta)$ of the scalar $t$
is replaced by the values
$S_k(\mat{x},\sigma,\theta)$ indexed by $k=1,\ldots, K$.
This is interesting for three reasons: 
Firstly, 
the model becomes inherently local, because the filters $G_k$
that compute the values $S_k$ are now centred at the same point
$\mat{x}$.
Secondly, 
the filters $G_k$ are symmetric
or antisymmetric about the point $\mat{x}$, and resemble the
Gabor functions used in the energy model.
Thirdly, 
the values $S_k$ can be obtained from a linear transformation of the 
$K$-th order local jet at $\mat{x}$, and so this scheme is compatible
with the scale space theory described above.

To be more specific, it will be shown that the first-order structure 
in the neighbourhood $\mat{x}+t\mat{v}$, as in (\ref{eqn:complex}), can be estimated from a 
linear combination of the directional derivatives $S_k$,
\begin{equation}
S_1(\mat{x}+t\mat{v},\sigma,\theta) \approx
\sum_{k=1}^K
P_k(t)\mspace{2mu} 
S_k(\mat{x},\sigma,\theta)
\label{eqn:approx-simple}
\end{equation}
where the functions $P_k(t)$ are fixed polynomials. 
This approximation will then be substituted into the right-hand side of 
(\ref{eqn:complex}).
It will be shown in section \ref{sec:functional} that the approximation 
(\ref{eqn:approx-simple}) can be motivated by a Maclaurin expansion
in powers of $t$.
This can also be interpreted, as shown in figure~\ref{fig:construct}, as the synthesis 
of spatially offset filters, using the Gaussian derivatives as a basis.
A matrix formulation of this model will be given in section \ref{sec:matrix}.
An optimal (and image-independent) construction of the polynomials $P_k(t)$
will be given in section \ref{sec:unconstrained}.
The case in which the derivatives on the right-hand side of 
(\ref{eqn:approx-simple})
are in another direction $\phi \ne \theta$, is treated in section 
\ref{sec:constrained}.

\subsection{Contributions \& Organization}
\label{sec:overview}
\noindent
The model presented in this paper is quite different from the previous approaches,
as explained above. 
The main contribution is a `differential' model of the complex cell, which is 
exactly steerable, and which fits naturally into the geometric approach to image analysis 
\citep{koenderink-1987,koenderink-1990}. 
This shows that it is possible to analyze the local image geometry, and to 
obtain a shift-invariant response, using a common set of filters.

The body of the paper is organized as follows.
The new model is developed in section \ref{sec:offset}, using linear algebra and
least-squares optimization.
The accuracy of the estimated filters is evaluated in section~\ref{sec:evaluation}.
This section also derives the exact response of the ideal filters to several 
basic stimuli. Some preliminary experiments on natural images are reported.
The biological interpretation of the model, and its predictions, are discussed
in section \ref{sec:discussion}. Future directions, and the relationship of the 
model to scale-space theory \citep{koenderink-1984}, are also discussed.

\section{Differential Model}
\label{sec:offset}

\noindent
The following notation will be used here. Matrices and vectors
are written in bold, e.g.\ $\mat{M}$, $\mat{v}$, where $\mat{M}^\tp$ is the
transpose, and $\mat{M}^+$ is the Moore-Penrose inverse \citep{press-1992}.  
The convolution of functions is 
$F(x)\star G(x) = \int_{-\infty}^\infty F(x-y)\, G(y)\, \mathrm{d}y$. Some 
properties of the Gaussian derivatives $G_k(x,\sigma)$
will now be reviewed.
There is no particular spatial scale
at which a natural image should be analyzed. It is therefore desirable to 
represent the image in a \emph{scale space}, so that a range of resolutions 
can be considered \citep{koenderink-1987}.
The preferred way to do this is by convolution with a Gaussian kernel.
It follows that the structure of the image, at a given scale, can be 
analyzed via the spatial derivatives of the corresponding Gaussian.
The \mbox{$k$-th} order derivatives of a \oned\ Gaussian,
$G_k = \mathrm{d}^k\!/\mathrm{d}x^k\, G_0$ can be expressed as
\begin{align}
G_k(x,\sigma) &=
\left(
\frac{-1}{\sigma\sqrt{2}}
\right)^{\!k}
H_k\left(
\frac{x}{\sigma\sqrt{2}}
\right)\,
G_0(x,\sigma) \label{eqn:deriv} \\[1ex]
G_0(x,\sigma) &= 
\exp\left(\frac{-x^2}{2\sigma^2}\right)
\label{eqn:gderivs}
\end{align}
where $G_0(x,\sigma)$ is the original Gaussian, $k$ is a positive integer, 
and $H_k(x)$ is the \mbox{$k$-th} Hermite polynomial.
The first seven Gaussian derivatives are shown in column one of figure
\ref{fig:construct}.
It will also be useful to introduce two normalizations of the Gaussian
derivatives. 
\begin{equation}
G_k^0(x,\sigma) = \frac{1}{\sigma\sqrt{2\pi}}\, G_k(x,\sigma) 
\eqand
G_k^1(x,\sigma) = \frac{1}{2}\, G_k(x,\sigma) 
\label{eqn:gauss-normalized}
\end{equation}
which are defined so that $\int |G_k^k(x)| \mathrm{d}x = 1$. In particular,
$G_0^0$ and $G_1^1$ are the \mbox{$L^1$-normalized} blurring and differentiating
filters, respectively. This superscript notation will not be used unless a
particular normalization is important (e.g.\ in sec.~\ref{sec:response}). 

The two-dimensional Gaussian derivative, in direction $\theta$ with $\mat{x}=(x,y)^\tp$,
will be written $G_k(\mat{x},\sigma,\theta) = (\mat{v}\cdot\nabla)^k G_0(\mat{x},\sigma)$,
as in section \ref{sec:present}.
Two special properties of these filters should be noted.
Firstly, the filter $G_k(\mat{x},\sigma,\theta)$ is \emph{separable} in the local 
coordinate-system that is defined by the direction of differentiation.
This means that the \twod\ filter can be obtained from the product of \oned\ filters
$G_k(x_\theta,\sigma)$ and $G_0(y_\theta,\sigma)$.
Secondly, the 
Gaussian derivatives are \emph{steerable}, meaning that $G_k(\mat{x},\sigma,\theta)$ 
can be obtained from a linear combination of derivatives in other directions, 
$G_k(\mat{x},\sigma,\phi_j)$, where $j=1,\ldots, k+1$. These facts make it possible
to analyze a multidimensional filter, in many cases, in terms of \oned\ functions
\citep{freeman-1991}.

The first derivative, $G_1$, will be used as the basic model of a complex 
subunit (which is also a simple cell receptive field).
This choice is motivated by two observations. 
Firstly, it is well established that gradient filters can be used to detect 
edges, as well as more complex image-features \citep{canny-1986,harris-1988}.
Secondly, $G_1$ is the first zero-mean filter in the local-jet representation 
of the image, which is physiologically and 
mathematically convenient. The extension to higher-order subunits is 
straightforward, as discussed in section \ref{sec:extensions}. 

\subsection{Filter Arrays}
\label{sec:arrays}

\noindent
This section will put the system of simple cells, introduced in section
\ref{sec:present}, into a standard signal processing framework. This will be done
in \oned, in order to simplify the notation. The extension to \twod\ is 
straightforward. 
The \oned\ version of the simple cell response (\ref{eqn:simple}) is 
$S_k(u,\sigma) = \int_{-\infty}^\infty G_k(u-x,\sigma) S(x) \mathrm{d}x$. 
If $k=1$, and the filter $G_1$ is offset by an amount $t$, then the convolution 
can be expressed as
\begin{equation}
\begin{aligned}
S_1(u-t,\sigma) &= 
\int_{-\infty}^\infty G_1(u-t-x,\sigma)\, S(x) \, \mathrm{d}x \\
&= -\int_{-\infty}^\infty 
G_1(x-t,\sigma)\, S(x-u) \, \mathrm{d}x.
\end{aligned}
\label{eqn:corrl}
\end{equation}
Here the antisymmetry of $G_1(x,\sigma)$ has been used to show
that the result is equal to the negative \emph{correlation} of the filter
and signal.
It is evident that if $t$ could be kept equal to $u$, then the signal shift
would have no effect on the result.
The prototypical stimulus for $G_1$ is the step-edge $S(x)=\mathrm{sgn}(x)$. 
In this case the filter and signal are anti-correlated,
and so the final response (\ref{eqn:corrl}) is non-negative.

\subsection{Functional Representation}
\label{sec:functional}

\noindent
It was established in the section \ref{sec:arrays} that the response $S_1(u-t,\sigma)$
can be constructed from the offset filters $G_1(x-t,\sigma)$. This means that 
the desired approximation (\ref{eqn:approx-simple}) can be treated as a filter-design
problem. The following notation will be adopted for the offset filters:
\begin{equation}
\begin{aligned}
F(0,x) &= G_1(x,\sigma)\\
F(t,x) &\approx  G_1(x-t,\sigma)
\end{aligned}
\label{eqn:base}
\end{equation}
which also depend on the spatial scale and derivative order, but it
will not be necessary to make this explicit in the notation. It will suffice
to analyze a single filter which, without loss of generality, is located 
at the origin $\mat{x}=(0,0)^\tp$ of the spatial coordinate-system.
The \emph{linear response} of this filter is defined in relation to (\ref{eqn:corrl}) as
\begin{equation}
R(t,u) = 
-\int_{-\infty}^\infty F(t,x)\, S(x-u) \, \mathrm{d}x.
\label{eqn:resp}
\end{equation}
The complex response at $\mat{x}=(0,0)^\tp$, with reference to (\ref{eqn:complex}), 
can now be expressed as a function of the signal translation $u$;
\begin{equation}
C(u) =
\max_t\, 
\bigl|
R(t,u)
\bigr| 
\eqwhere  |t| \le \rho.
\label{eqn:max}
\end{equation}
The actual value of $u$, in general, has no particular
significance. It will be more important to consider the response $R(t,u)$ as
$u$ changes. In particular, suppose that $|R(t,u)|$ is high at the stimulus 
position $u=u_0$. If the response is insensitive to slight translation of the signal,
then $\partial^2 C \big/ \partial u^2 \approx 0$ at $u_0$.

The approximation problem in (\ref{eqn:base}) will now be addressed.
The filter $F(t,x)$ can be defined in relation to the Maclaurin 
expansion of $G_1(x-t,\sigma)$ with respect to the offset $t$, as indicated
in (\ref{eqn:base}). 
If image-derivatives up to order $K$ are available, then the approximation is
\begin{equation}
F(t,x) = \sum_{k=0}^{K-1} \frac{(-t)^k}{k!}\, G_{k+1}(x,\sigma) 
\label{eqn:deriv-series}
\end{equation}
The key observation is that the filters $G_{k+1}(x,\sigma)$ 
in (\ref{eqn:deriv-series}) are precisely those that compute the local jet coefficients,
of order $1,\ldots, K$, at the point $x=0$. In other words, the family of shifted filters 
$F(t,x)$ has been obtained from the family of \emph{non-shifted} derivatives $G_k(x,\sigma)$.
It can be seen from (\ref{eqn:deriv},\ref{eqn:gderivs}) and (\ref{eqn:deriv-series}) that 
the estimated function 
$F(t,x)$ decreases to zero for large $|x|$, owing to
the exponential tails of $G_0$.

\begin{figure}[!ht]
\begin{center}
\includegraphics{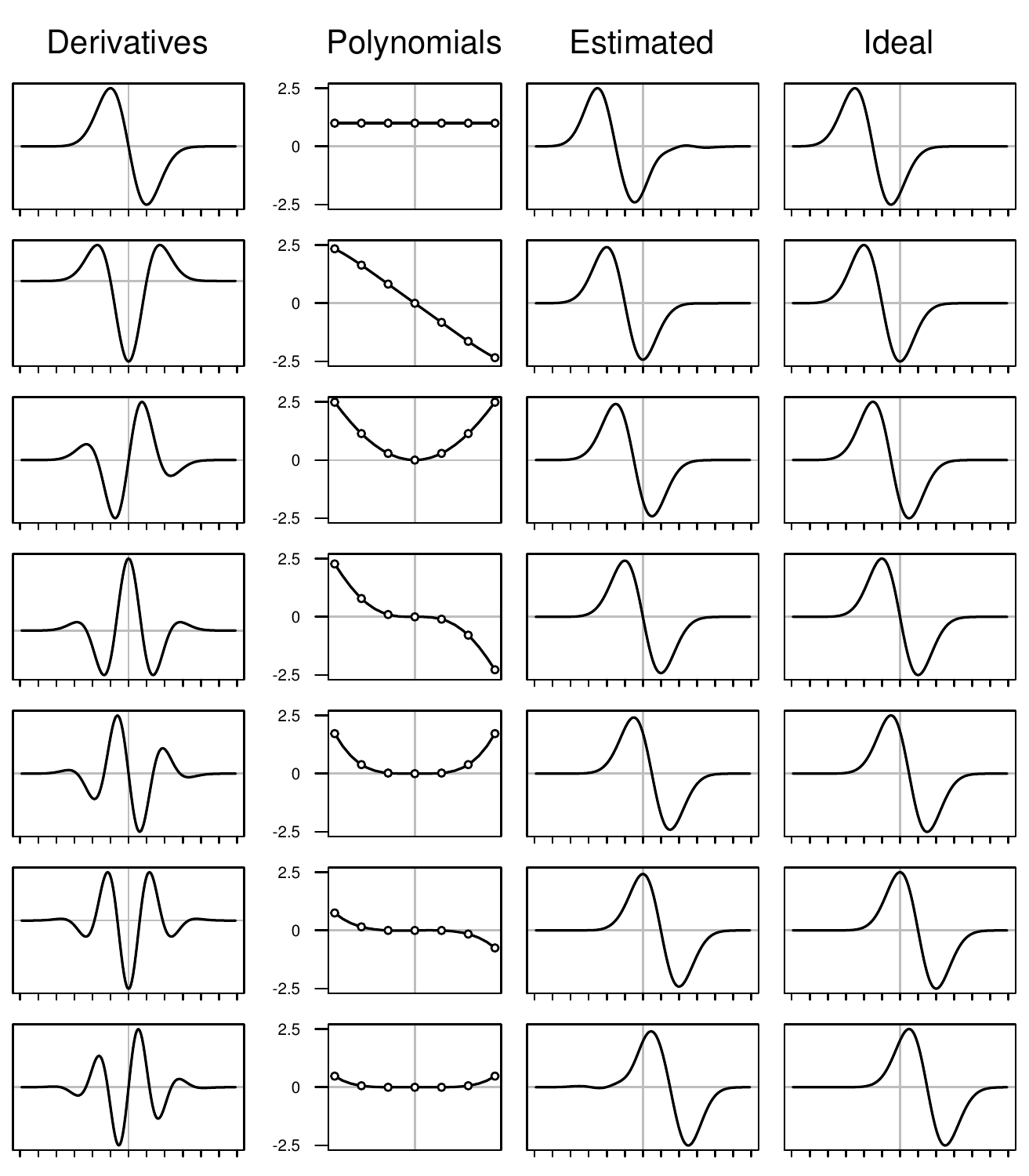}
\caption{{Construction of offset filters}. \textbf{Column 1:} The Gaussian 
derivatives $G_k(x,\sigma)$, scaled for display, of orders $1,\ldots,K$, where
$K=7$. 
\textbf{Column 2:} The corresponding polynomial interpolation functions $P_k(t)$,
of order $K-1$. Note that $P_k(t)$ resembles the monomial $t^{(k-1)}$.
\textbf{Column 3:} Estimated filters, $F(t_j,x)$ which are offset versions of $G_1(x,\sigma)$. 
\textbf{Column 4:} Ideal filters $F_\star(t_j,x)$.
The synthesis equation is $F(t_j,x) = \sum_k P_k(t_j) G_k(x,\sigma)$, where each 
weight $P_k(t_j)$ corresponds to the $j$-th dot on the $k$-th polynomial in 
column two.
}
\label{fig:construct}
\end{center}
\end{figure}

The definition (\ref{eqn:deriv-series}) is usable in practice, as will 
be shown in section \ref{sec:error}.
There are, however, two difficulties with the scheme described above.
Firstly, although $F(t,x)$ is an approximation of $G_1(x-t,\sigma)$,
the nature of this approximation (a polynomial with the given derivatives) 
may be inappropriate. Secondly, as expected, the approximation 
(\ref{eqn:deriv-series}) is not well-behaved for large $|t|$.
Both of these problems can be addressed by 
replacing the Maclaurin series with a more flexible construction of
$F(t,x)$. This is done by substituting a general polynomial $P_k(t)$ in place of 
each monomial $(-t)^k/k!$ in (\ref{eqn:deriv-series}), leading to
\begin{equation}
F(t,x) = \sum_{k=1}^K P_k(t)\, G_k(x,\sigma).
\label{eqn:filt}
\end{equation}
The $K$ polynomials $P_k(t)$ are constructed from standard monomial basis
functions $t^j$ and coefficients $c_{jk}$. The order of each 
polynomial will be $K-1$, for consistency with the original series
approximation (\ref{eqn:deriv-series}). It follows that the polynomials 
are
\begin{equation}
P_k(t) =
\sum_{j=0}^{K-1}\, t^j c_{jk} \eqwhere
1 < k \le K.
\label{eqn:poly}
\end{equation}
The problem has now been altered to that of finding $K^2$ appropriate
coefficients $c_{jk}$.
This will be treated, in sections \ref{sec:unconstrained} and \ref{sec:constrained},
as the optimization of
\[
\arg\min_{\mat{C}}
\iint \, \bigl|F(t,x)-G_1(x-t,\sigma)\bigr|^2\mspace{3mu}\mathrm{d}x\,\mathrm{d}t
\]
where $F(t,x)$ is the family of filters defined in (\ref{eqn:filt}), and $\mat{C}$
is the matrix of coefficients $c_{jk}$. This optimization scheme generalizes 
immediately to filters in any number of dimensions.
The simple Maclaurin scheme (\ref{eqn:deriv-series}) remains a useful model, 
because the optimal polynomials are, in practice, close to the original 
monomials $P_k(t)\approx t^{(k-1)}$, as can be seen in figure~\ref{fig:construct}.

It is important to note that, once the coefficients $c_{jk}$ have been estimated,
the location of the synthetic filter $F(t,x)$ can be varied \emph{continuously} with 
respect to the offset $t$. Any set of translated filters $F(t_i,x)$ can
be obtained, provided $|t_i|\le \rho$ for $i=1,\ldots M$, by re-sampling the
monomial basis functions as $t_i^j$ and, then repeating (\ref{eqn:filt} \& \ref{eqn:poly}). 
Furthermore, the principle of
\emph{shiftability} states that the convolution $f(x-t)\star s(x)$ can be represented 
in a finite basis $f(x-t_i)$, provided that the filter $f$ is bandlimited 
\citep{simoncelli-1992,perona-1995}. The filters $F(t,x)$ are not bandlimited, but they 
do decay exponentially, as the Fourier transforms have a Gaussian factor. 
This means, in practice, that the linear response $R(t,u)$ in (\ref{eqn:resp}) 
can be represented by a suitable discretization $R(t_i,u)$, where the shift-resolution 
$\Delta t = 2\rho / (M-1)$ 
can be chosen to achieve any desired accuracy.

\subsection{Matrix Representation}
\label{sec:matrix}

\noindent
It will be convenient to represent the filter construction in terms
of matrices. This results in a compact formulation, and prepares
for the least-squares estimation procedure that will be introduced in 
section \ref{sec:unconstrained}. Suppose that $M$ filters, each of length
$N$ are to be constructed, and that the highest available derivative 
is of order $K$. Each filter will be represented as a row-vector,
so that the collection of offset filters forms an $M\times N$ matrix $\mat{F}$.
Note that this representation applies in any number of dimensions, provided 
that the positions of the filter-samples are consistently identified with the 
column-indices of $\mat{F}$.

The columns of another matrix $\mat{P}$ will contain the $K$ polynomials $P_k(t)$ 
from equation (\ref{eqn:poly}).
These polynomials must be sampled at $M$ points $t_i$, hence $\mat{P}$ has
dimensions $M\times K$. Let the sampled monomial basis functions $t_i^j$
be the columns of the matrix $\mat{B}$, which must therefore have the same dimensions,
$M\times K$. 
The monomials are weighted by
the $K\times K$ matrix of coefficients $\mat{C}$, such that
\begin{equation}
\mat{P} = \mat{B} \mat{C}
\label{eqn:poly-mat}
\end{equation}
where column $k$ of $\mat{C}$ contains the coefficients of 
$P_k$.
Let each row of the $K\times N$ matrix $\mat{G}$ contain the sampled Gaussian 
derivative $G_k(x_i,\sigma)$. Each offset filter
should be a linear combination of the Gaussian derivatives, constructed from
the polynomials $P_k$.  It follows from
(\ref{eqn:filt}) that
\begin{equation}
\mat{F} = \mat{P}\mat{G}.
\label{eqn:filt-mat}
\end{equation}
Let the column-vector $\mat{s}$ contain the sampled signal, 
$
\mat{s} = \bigl(S(x_1),\ldots,S(x_N)\bigr)^{\!\tp}
$. 
This means that the response-vector 
$
\mat{r} =
\bigl(R(t_1,u),\ldots,R(t_M,u)\bigr)^{\!\tp}
$
is obtained according to (\ref{eqn:resp} \& \ref{eqn:filt-mat}) as
\begin{equation}
\begin{aligned}
\mat{r} &= -\mat{F}\mat{s}\\[.25ex]
&= -\mat{P} \mat{G} \mat{s}.
\end{aligned}
\label{eqn:resp-vec}
\end{equation}
This clearly shows that the response $\mat{r}$ is simply a 
linear transformation $\mat{P}$ of the \mbox{$K$-th}
order Gaussian jet,
$
\mat{G}\mat{s} = \bigl(S_1(x,\sigma),\ldots,S_K(x,\sigma)\bigr)^{\!\tp}
$.
The implication is that the filter-bank $\mat{F}$ need not be explicitly 
constructed; rather, the response $\mat{r}$ is computed directly from the 
$K$ image derivatives $S_k$ at $\mat{x}$. 

\begin{figure}[!ht]
\begin{center}
\includegraphics{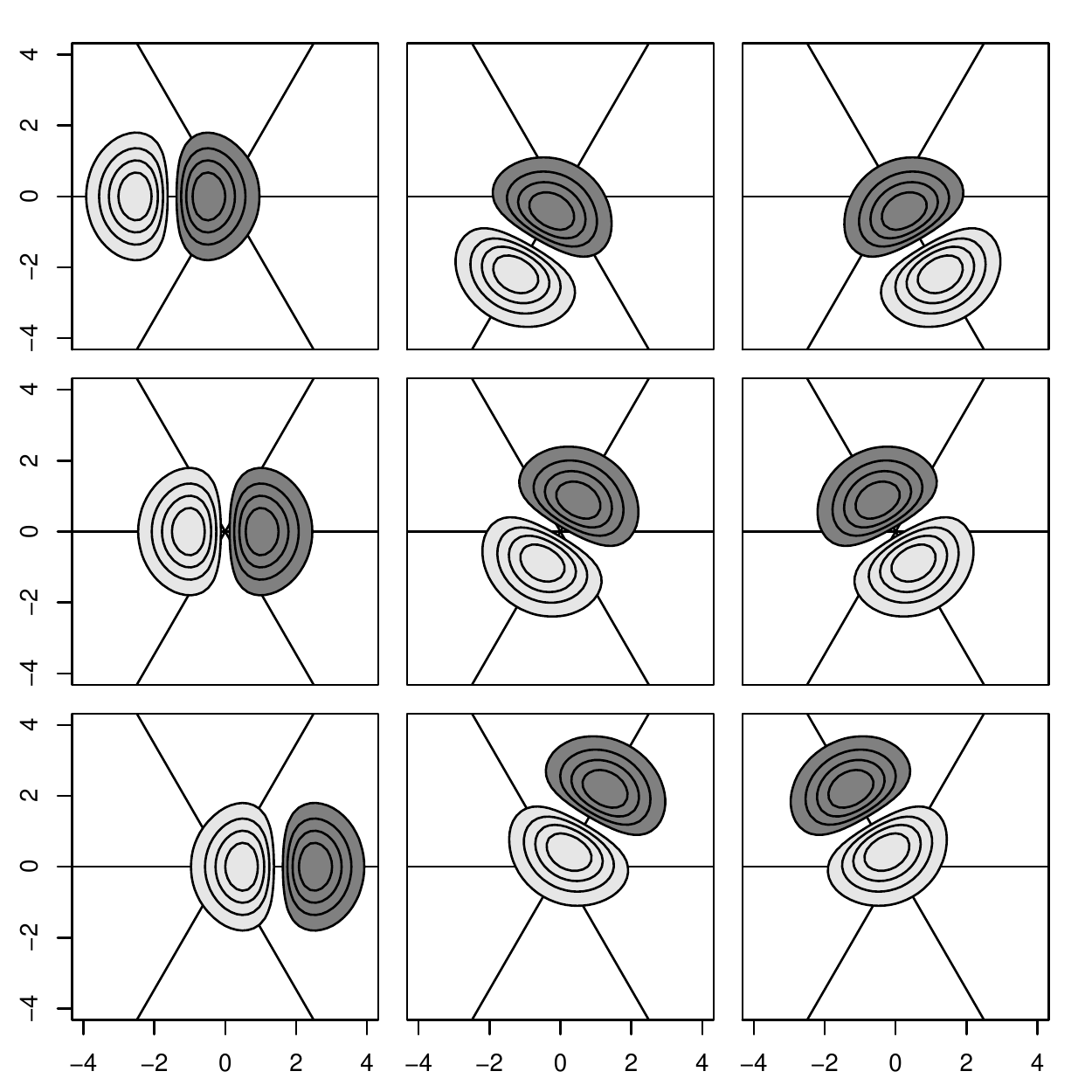}
\caption{{Synthesis of orientation-tuned subunits}. The nine 
filters were synthesized from eight oriented derivatives 
centred at the origin (with of $\sigma=1$). The steered-additive solution of section~\ref{sec:addsteered}
was used. \textbf{Middle row:} The $G_1$ filter is steered to the
axes of a hexagonal lattice; $\theta = 0^\circ\!, 60^\circ\!, 120^\circ$. 
\textbf{Top and bottom rows:} Offset filters, synthesized at
shifts of $t=\pm \rho$ in the corresponding 
directions (with $\rho=1.5$). Note that each column can be interpreted as three subunits of
an orientation-tuned complex cell. The filter amplitudes are scaled to the
range $[-1,1]$, with contour lines separated by increments of 0.2 units.}
\label{fig:hexfilt}
\end{center}
\end{figure}

\subsection{Unconstrained Estimation}
\label{sec:unconstrained}

\noindent
It will now be shown that the $K^2$ unknown 
coefficients, contained in the matrix $\mat{C}$, can be obtained by standard
least-squares methods. It should be emphasized that this is a \emph{filter-design}
problem; the matrix $\mat{P}$ is fixed for all signals, and the response $\mat{r}$
is obtained according to (\ref{eqn:resp-vec}).

Let the $M\times N$ matrix $\mat{F}_\star$ contain the \emph{true} derivative
filters, such that the 
\mbox{$ij$-th} element is
$G_1(x_j-t_i,\sigma)$. The approximation $\mat{F}\approx\mat{F}_\star$ can be
expressed, according to (\ref{eqn:poly-mat} \& \ref{eqn:filt-mat}), as the product
\begin{equation}
\mat{F}_\star
\approx 
\mat{B} \mat{C} \mat{G},
\quad\text{of which}\quad
\mat{C} = \mat{B}^+ \mat{F}_\star \mat{G}^+
\label{eqn:approx}
\end{equation}
is the solution in the least-squares sense.
This formulation requires two Moore-Penrose inverses, which can be computed from 
the singular-value decompositions of the monomial basis and Gaussian derivative 
matrices $\mat{B}$ and $\mat{G}$, respectively. It is, however, more efficient to 
solve this problem using \emph{QR} decompositions, as follows. There are, in practice, 
more offsets than derivative filters $M > K$, as well as
more spatial samples than derivative filters $N > K$.
The basis matrix has full column-rank $K$, and can be factored as 
$\mat{B}=\mat{Q}_B\mat{R}_B$. 
The derivative matrix has full row-rank $K$, and so its transpose can be factored as
$\mat{G}^\tp=\mat{Q}_G\mat{R}_G$.
It follows that the solution (\ref{eqn:approx}) can be obtained via
$\mat{B}^+ = \mat{R}_B^{-1}\mat{Q}_B^\top$
and
$\mat{G}^+ = \mat{Q}_G \mat{R}_G^{-\!\top}$.

\subsection{Constrained Estimation}
\label{sec:constrained}

\noindent
The least-squares construction of the filters $\mat{F}$ was described in the 
preceding section. The method is quite usable, but has two shortcomings. Firstly,
if one of the shifts $t_i$ is zero, then
$F(0,x) \approx G_1(x,\sigma)$, but it would  be preferable to make this an \emph{exact}
equality, so that the original filter is returned as in (\ref{eqn:base}).
The second shortcoming of the method in \ref{sec:unconstrained} is that, in two or more dimensions, 
the \emph{orientation} of the derivative filters in the basis-set $\mat{G}$ may not
match that of the target-set $\mat{F}_\star$. Both of these problems will be 
solved below.

\subsubsection{Additive Solution}
\label{sec:additive}

The requirement $F(0,x) = G_1(x,\sigma)$ is satisfied by an 
\emph{additive} model, in which the polynomial $P_1(t)$ that 
weights $G_1(x,\sigma)$ is always unity, and all other polynomials
pass through zero when $t=0$.
This implies the following partitioning of the derivative, monomial and coefficient
matrices:
\begin{equation}
\mat{G} =
\begin{pmatrix}
\mat{g}_1^\tp \\[.75ex]
\mat{G}_\Delta
\end{pmatrix}
\quad
\mat{B} =
\begin{pmatrix}
\mat{1} \mspace{10mu} \mat{B}_\Delta
\end{pmatrix}
\quad
\mat{C} =
\begin{pmatrix}
1       & \mat{0}^\tp \\[.5ex]
\mat{0} & \mat{C}_\Delta
\end{pmatrix}
\end{equation}
where $\mat{1}$ is the column-vector of $M$ ones, and $\mat{0}$ is the 
column-vector of $(K-1)$ zeros.
The $1\times N$ vector $\mat{g}_1^\tp$ contains the first derivative filter $G_1(x,\sigma)$, while
the $(K-1)\times N$ matrix $\mat{G}_\Delta$ contains the higher-order filters.
The columns of the $M\times (K-1)$ matrix $\mat{B}_\Delta$ contain the sampled monomials,
\emph{excluding} the constant vector $\mat{1}$. The unknown matrix $\mat{C}$ will 
be recovered in the form indicated, where $\mat{C}_\Delta$ has dimensions $(K-1)\times(K-1)$. 

The product of $\mat{B}$ and $\mat{C}$, as in (\ref{eqn:poly-mat}), now
gives $\mat{P} = (\mat{1}\mspace{10mu} \mat{P}_\Delta)$, where the columns of
$\mat{P}_\Delta=\mat{B}_\Delta\mat{C}_\Delta$ are polynomials \emph{without} constant terms.
It follows that the product $\mat{P}\mat{G}$, as in (\ref{eqn:filt-mat}), gives the 
additive approximation 
\begin{equation}
\mat{F}_\star
\approx
\mat{1}\mat{g}_1^\tp + \mat{B}_\Delta\mat{C}_\Delta\mat{G}_\Delta
\label{eqn:additive}
\end{equation}
where
$\mat{1}\mat{g}_1^\tp$ is the rank-one matrix containing $M$ identical rows $\mat{g}_1^\tp$.
Note that if the \mbox{$i$-th} row of $\mat{F}_\star$ corresponds to $t=0$, 
then the \mbox{$i$-th} row of $\mat{B}_\Delta\mat{C}_\Delta$ must be zero, this being
the evaluation of the polynomials $\sum_{j=1}^{K-1} t^j c_{jk}$ at $t=0$. It follows that the 
\mbox{$i$-th} row of $\mat{F}_\star$ is exactly recovered from (\ref{eqn:additive}) 
as 
$\mat{g}_1^\tp$, and so the constraint $F(0,x) = G_1(x,\sigma)$ has been imposed.

The unknown coefficients $\mat{C}_\Delta$ are recovered by subtracting 
$\mat{1}\mat{g}_1^\tp$ from $\mat{F}_\star$, and then proceeding by analogy with
(\ref{eqn:approx}). This leads to 
\[
\mat{C}_\Delta = \mat{B}_\Delta^+ 
\bigl(\mat{F}_\star-\mat{1}\mat{g}_1^\tp\bigr) 
\mat{G}_\Delta^+
\]
where the matrices $\mat{B}_\Delta^+$ and $\mat{G}_\Delta^{+\!\top}$ can be obtained from 
the \emph{QR} factorizations of $\mat{B}_\Delta$ and $\mat{G}_\Delta$, as before. 

\subsubsection{Steered Solution}

\label{sec:steered}

In two (or more) dimensions, it is assumed that the desired filters 
$G_1(x-t,\theta,\sigma)$ have a common orientation, where $\mat{v}=(\cos\theta,\sin\theta)^\tp$ 
is the direction of the derivative in \twod. This leads to invariance 
with respect to translations of the signal in the given direction. The basis filters 
$G_k(x,\sigma,\theta)$, however, will typically have a range of orientations 
$\phi_\ell \ne \theta$. 
This problem can be solved as follows.

Recall from section \ref{sec:offset} that the \mbox{$k$-th}
order Gaussian derivative is \emph{steerable} with a basis of size $k+1$. Now suppose
that row $k$ of the matrix $\mat{G}$ is replaced by $k+1$ rows, containing
sampled filters $G_k(x,\phi_\ell,\sigma)$ at $\ell=1,\ldots,k+1$ distinct orientations $\phi_\ell$. 
The enlarged matrix $\mat{G}_\phi$ now has dimensions $M_K \times N$, where
\begin{equation}
M_K = \sum_{k=1}^K (k+1)
       = \tfrac{1}{2} K(K+3).
\label{eqn:steernum}
\end{equation}
It follows that there is a $K\times M_K$ `steering' matrix $\mat{D}$ such that 
$\mat{G} = \mat{D}\mat{G}_\phi$ is exactly
the $K \times N$ matrix  of derivatives at the desired orientation.
Moreover, if the approach of section \ref{sec:unconstrained} is applied to the $M_K \times N$ 
matrix $\mat{G}_\phi$, then a solution
\begin{equation}
\mat{F}=\mat{B}\mat{C}_\phi\mat{G}_\phi =
\mat{B} \mat{C} \mat{G}
\end{equation}
will be obtained.
It follows that the two coefficient matrices are related by
$\mat{C}_\phi = \mat{C}\mat{D}$. In summary, if the matrix
$\mat{G}$ contains a \emph{sufficient} number $M_K$ of differently oriented filters, 
then a set of translated filters $\mat{F}$ can be approximated in any 
common orientation $\theta$. There is no change to the algorithm described in section
\ref{sec:unconstrained}.
It should, however, be noted that (\ref{eqn:steernum}) shows a trade-off between 
translation invariance and steerability. Larger translation invariance
requires more derivatives, but these become increasingly difficult to steer.

\subsubsection{Additive Steered Solution}
\label{sec:addsteered}

The steered solution, as described in the previous section, will not
automatically be additive, in the sense of (\ref{eqn:additive}).
This problem will be solved, with reference to section~\ref{sec:additive},
by putting an explicitly steered filter $\mat{g}_\theta^\tp$ in place of
$\mat{g}_1^\tp$.
The first derivative
can be steered with respect to a basis of filters at distinct
orientations $\phi_1$ and $\phi_2$ (these would be the first two 
rows of the $M_K\times N$ matrix $\mat{G}_\phi$). The standard steering
equation \citep{freeman-1991} can be simplified in this case to
\[
\begin{pmatrix}
\cos\theta\\
\sin\theta 
\end{pmatrix} =
\begin{pmatrix}
\cos\phi_1 & \cos\phi_2\\
\sin\phi_1 & \sin\phi_2
\end{pmatrix}
\begin{pmatrix}
p_1\\
p_2
\end{pmatrix}
\]
where $\theta$, $\phi_1$ and $\phi_2$ are known angles. This system can be
solved exactly for the unknown coefficients $p_1$ and $p_2$, resulting in
\begin{equation}
\begin{gathered}
p_1 = \sin(\phi_2-\theta) / \delta \eqand
p_2 = \sin(\theta-\phi_1) / \delta \\
\eqwhere \delta = \sin(\phi_2 - \phi_1).
\end{gathered}
\end{equation}
It may be noted that if $\phi_1=0$ and $\phi_2=\pi/2$, then the solution reduces
to the usual coefficients $p_1=\cos\theta$ and $p_2=\sin\theta$ for the construction
of the directional derivative from $\mathrm{d}/\mathrm{d}x$ and 
$\mathrm{d}/\mathrm{d}y$.
The additive steered approximation can now be defined, using the new filter
$G_1(\mat{x},\theta,\sigma)$, as
\begin{equation}
\mat{F}_\star \approx 
\mat{1}\mat{g}_\theta^\tp +
\mat{B}_\Delta \mat{C}_{\phi\Delta} \mat{G}_{\phi\Delta} \eqwhere
\mat{g}_\theta^\tp = p_1\mat{g}_{\phi_1}^\tp + p_2\mat{g}_{\phi_2}^\tp.
\label{eqn:addsteer}
\end{equation}
This system can be solved in the same way as (\ref{eqn:additive}). Note that the 
higher-order filters in $\mat{G}_{\phi\Delta}$ will be implicitly steered,
as described in section \ref{sec:steered}.

\section{Evaluation}
\label{sec:evaluation}

\noindent
Two issues are addressed in this evaluation, as follows.
\emph{Approximation}:
The accuracy of the least-squares algorithms from sections \ref{sec:unconstrained} 
and \ref{sec:constrained} is established in section~\ref{sec:error}.
\emph{Characterization}: The response of the underlying model from section~\ref{sec:present}
to basic stimuli, as well as to natural images, is analyzed in 
sections~\ref{sec:response} and \ref{sec:images} respectively. 
Note that the issues of approximation and characterization are addressed separately, 
in order to avoid mixing different sources of error. Hence section \ref{sec:error} 
will evaluate the approximate filters $F$, while sections \ref{sec:response} and 
\ref{sec:images} will analyze the ideal filters $F_\star$.

\subsection{Approximation Error}
\label{sec:error}
\noindent
The accuracy of the filter approximations will be evaluated in this
section, and it will be shown that the least-squares methods are 
superior to the original Maclaurin expansion.
The evaluation is based on the root mean-square (RMS) error
between the target and synthetic filters.

The accuracy of a given filter-synthesis method is determined by two variables;
the range of offsets, and the number of available derivatives (size of 
the basis). Better approximations can, in general, be obtained by reducing
the range of offsets and/or increasing the size of the basis. 
The range $\rho=1\sigma$ is the smallest that results in a unimodal impulse 
response, as will be shown in section \ref{sec:impulse}. It is therefore important
to analyze the corresponding approximation. In addition, the larger range
$\rho=1.5\sigma$ will be analyzed. This leaves the size of the basis (for which 
there is no prior preference) to be varied in each case.

\begin{figure}[!ht]
\begin{center}
\includegraphics{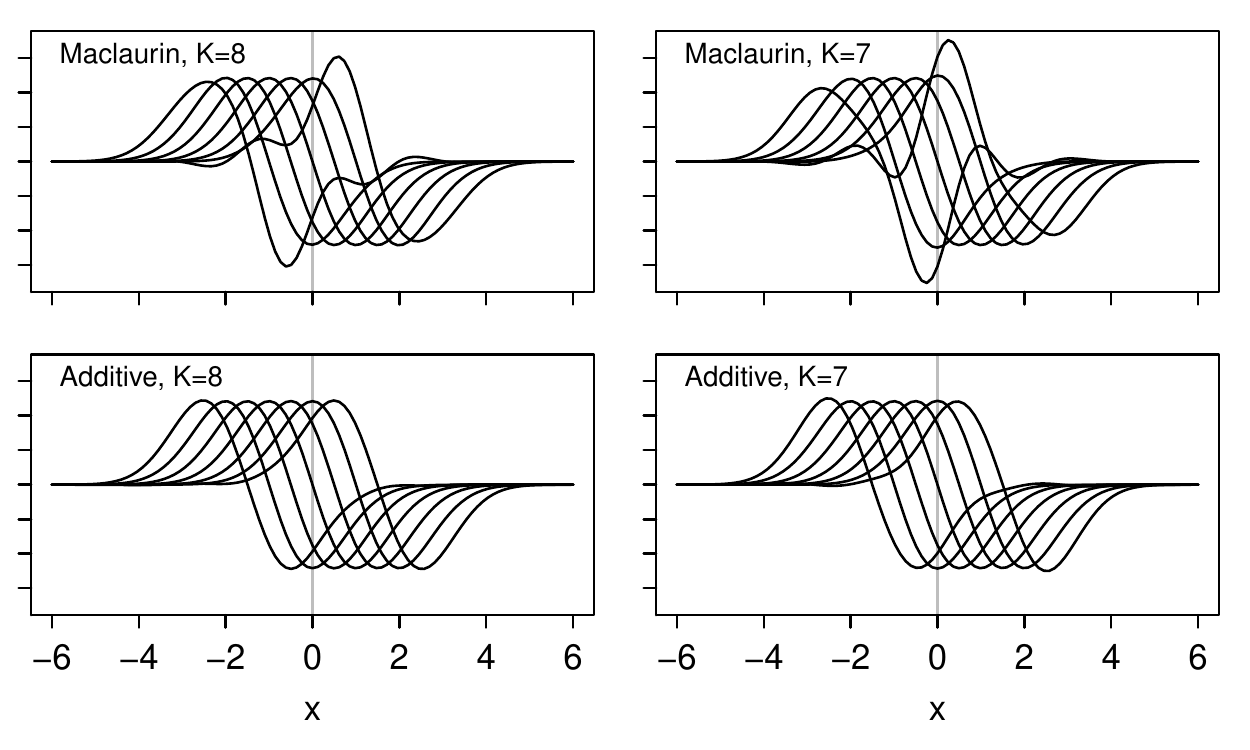}
\vspace*{-2ex}
\caption{{Filter deformation}. \textbf{Top left:} The eighth-order Maclaurin synthesis 
(\ref{eqn:deriv-series}) of filters $\sigma=1$, over a range $\rho=\pm1.5\sigma$ of offsets. Large 
errors are visible in the most extreme filters. 
\textbf{Top right:} The approximation is much worse if the order of 
the basis is reduced by one.
\textbf{Bottom left, right:} The least-squares approximation is much better,
even if the additivity constraint is enforced.}
\label{fig:deform}
\end{center}
\end{figure}

The method of evaluation is illustrated in figure~\ref{fig:deform}. It can be
seen that the furthest-offset filters begin to depart from the target shape.
The RMS difference between the ideal and approximate filters, for each test, 
was measured over 51 offsets $t_i$ in the range $\pm\rho$. Each filter was sampled
at 101 points $x_j$ in the range $\pm 6\sigma$, which contains the 
significantly nonzero part of all filters (see fig.~\ref{fig:deform}).

\begin{figure}[!ht]
\begin{center}
\includegraphics{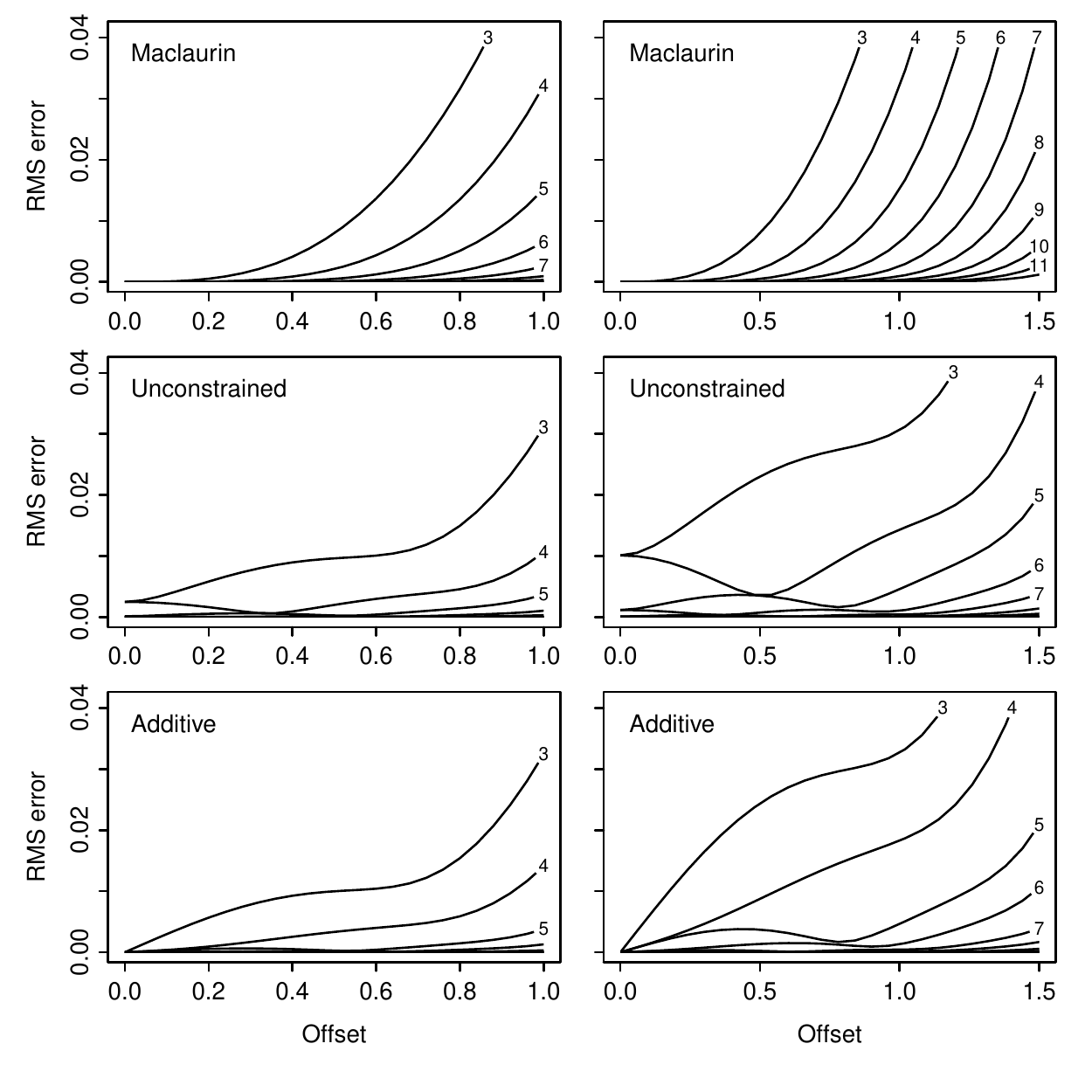}
\vspace*{-3ex}
\caption{{Error vs. offset}.
\textbf{Top row:} The Maclaurin error rises quickly as the target filter is offset
from the centre of the basis. Each line represents a different basis-size,
$k=3,\ldots,12$, as indicated. The left and right plots show ranges 
$\rho=1\sigma$, and $\rho=1.5\sigma$, respectively.
\textbf{Middle row:} The unconstrained least-squares approximation is much better, 
especially for high-order bases. \textbf{Bottom row:} The additive approximation is
also good, and ensures that the error is zero when there is no offset
(as in the Maclaurin case).}
\label{fig:errors}
\end{center}
\end{figure}

The RMS error, for basis sizes $K=3,\ldots,10$ is shown in fig.~\ref{fig:errors}.
The meaning of the RMS error, in terms of filter distortion, can be gauged with 
reference to fig.~\ref{fig:deform}. For example, it can be seen that the 
Maclaurin model with $K=8$ and $\rho=1.5$ is poor, and this corresponds to a point
around the middle of the RMS axis in fig.~\ref{fig:errors}.

In the case of the Maclaurin approximation (top row) it can be seen that the error increases 
rapidly and monotonically with respect to the offset. The pattern is more
complicated for the least-squares approximations, because the error has been 
minimized over an interval $\pm\rho$, which effectively truncates the basis
functions in $x$. Nonetheless, the lines corresponding to the different basis-sizes 
remain nested; they cannot cross, because increasing the size of the basis cannot 
make the approximation worse. It is, however, possible for the lines to meet. In 
particular, the unconstrained lines meet in pairs at $t=0$. This is because the 
target function at zero offset is anti-symmetric. It follows that the incorporation 
of a \emph{symmetric} basis function $G_{2k}$ cannot improve an existing approximation 
of order $2k-1$. In the case of the additive approximation, all lines meet
at $t=0$, where the error is zero by construction.

\subsection{Response to Basic Signals}
\label{sec:response}

\noindent
A number of basic signals will be introduced below, and the ideal
responses will be derived; further examples are given in \cite{hansard-2010}. 
The responses are `ideal' in the sense that the
error of the least-squares approximations (sec.\ \ref{sec:unconstrained}--\ref{sec:addsteered}) will
be ignored. This is primarily in order to obtain useful results, but
there are two further justifications. Firstly, it has been demonstrated in the 
preceding section
that the approximations are good, over an appropriate range $\rho$. Secondly, 
the approximation error can be made arbitrarily low, by using a large 
enough basis for the given range. 

Recall that the offset filters $F(t,x)$ are copies of the Gaussian derivative
$G_1(x,\sigma)$. It follows from (\ref{eqn:resp}) that the response function is covariant
to the shift $t$, in the sense that
\begin{equation}
R(t,u) = R(0,u-t).
\label{eqn:resp-trans}
\end{equation}
It therefore suffices to obtain the linear response for the case
$t=0$, as the other responses are simply translations of this function.
The linear response in this case is $R(0,u) = G_1^1(x,\sigma)\star S(x-u)$, 
by analogy with (\ref{eqn:corrl}). Note that the normalized filter has been used,
as defined in (\ref{eqn:gauss-normalized}). It follows that $R(0,u)$ can be
obtained by blurring the signal with the filter $G_0^1(x,\sigma)$, and then 
differentiating the result. The complex response $C(u)$ is given by the max 
operation (\ref{eqn:max}).
Evidently $C(u)$ is the \emph{upper envelope} of the 
family $|R(t,u)|$, but it is possible to be more precise than this. In 
particular, the shift-insensitivity of the model can be quantified by determining 
the intervals of $u$ over which $C(u)$ is constant, as described below.

The response $|R(0,u)|$ to a basic signal $S(x-u)$ can be either symmetric or antisymmetric, 
and either periodic or aperiodic. However, a common property of the
responses considered here is that the local maxima are all of equal height.
Let $|R(0,u^\star)| = R^\star$ be a local maximum, and suppose that $u$
is within range of this maximum, meaning that $|u-u^\star|\le\rho$. It follows that
$C(u) = R^\star$, because the maximum in (\ref{eqn:max}) will be found at 
$t=u-u^\star$, and $|R(u-u^\star,u)| = |R(0,u^\star)| = R^\star$ by (\ref{eqn:resp-trans}). 
An intuitive summary of this is that each local maximum $|R(0,u^\star)|$ generates a 
plateau $C(u^\star\pm\rho) = R^\star$ in the complex response. In order to make use of this
interpretation, the function $V(u)$ will be defined as the signed distance $u-u^\star$ 
to the nearest local maximum of $|R(0,u)|$. It follows that
\begin{equation}
C(u) = 
\begin{cases}
R^\star
&\text{if}\mspace{10mu} |V(u)| \le \rho\\
\underset{|t|\le\rho}{\max}\mspace{5mu} |R(t,u)|
&\text{otherwise}.
\end{cases}
\label{eqn:response}
\end{equation}
This explicitly identifies the intervals, $|V(u)|\le\rho$, over which $C(u)$ is
constant. Note that if $V(u)>\rho$ then the original definition (\ref{eqn:max})
is used. The functions $R(0,u)$ and $V(u)$, as well as the constant $R^\star$, will
now be derived for each of the basic signals. It should be emphasized that 
$V(u)$ and $R^\star$ are only used to \emph{characterize} the response; they
are \emph{not} part of the computational model.

\subsubsection{Impulse}
\label{sec:impulse}

\begin{figure}[!ht]
\begin{center}
\includegraphics{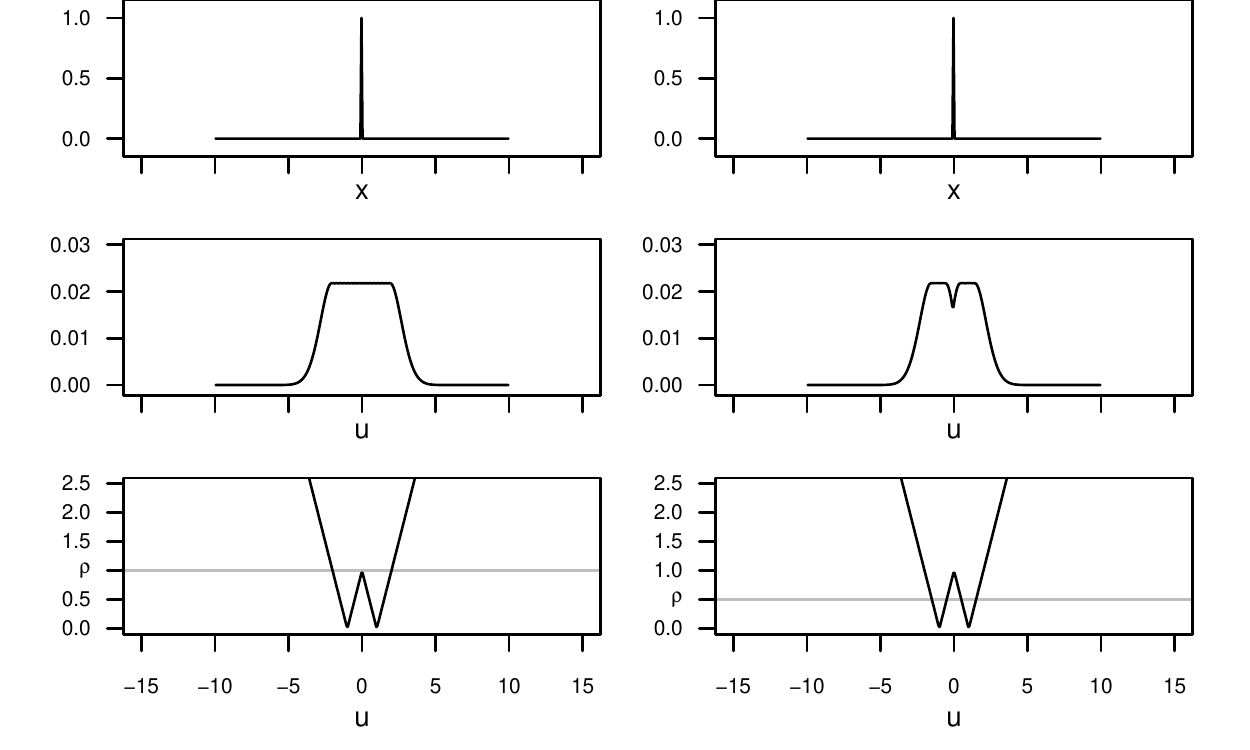}
\vspace*{-.1in}
\caption{{Impulse response}. \textbf{Left column:} the top plot shows the unit impulse, $S_\sigma(x)$,
as in (\ref{eqn:impulse-sig}). The middle plot shows the response $C(u)$. The bottom plot
shows the distance function $|V_\sigma(u)|$, as in (\ref{eqn:impulse-dist}), along with the value
of the maximum offset $\rho=1\sigma$. The response is constant, $C(u) = R^\star_\sigma$, when
$|V_\sigma(u)| \le \rho$. The critical case, $\sigma=1$, $\rho=1$ is plotted. 
\textbf{Right column:} as before, except $\rho=0.5\sigma$. The response becomes bimodal, which shows
the importance of the condition $\sigma \le \rho$.}
\label{fig:impulse}
\end{center}
\end{figure}

The first test signal to be considered is the unit impulse, which can be 
used to characterize the initial linear stage of the model. The impulse is
defined as
\begin{equation}
S_\sigma(x) = \delta(x)
\label{eqn:impulse-sig}
\end{equation}
where $\delta(x)$ is the Dirac distribution. It follows that the linear
response $G_1^1\star S_\sigma$ is just the original normalized derivative filter,
\begin{equation}
R_\sigma(0,u) = G_1^1(u,\sigma).
\label{eqn:impulse-resp}
\end{equation}
The maxima of the linear response can be found by differentiating 
$R_\sigma(0,u)$, and setting the result to zero. The derivative contains
a factor $\sigma^2-u^2$, and so the zeros are at $\pm\sigma$. The peak is
at $-\sigma$, and it follows that the maximum response is
\begin{equation}
R^\star_\sigma = G_1^1(-\sigma,\sigma).
\label{eqn:impulse-const}
\end{equation}
Both extrema become peaks in $|R_\sigma(0,u)|$, and the extent of the response 
plateau is determined by the minimum distance from these. 
The distance function for the impulse response can now be defined as
\begin{equation}
V_\sigma(u) = u - \mathrm{sgn}_+\mspace{-1mu}(u) \,\sigma,
\label{eqn:impulse-dist}
\end{equation}
where $\mathrm{sgn}_+$ is the sign-function with the convention $\mathrm{sgn}_+(0)=1$. 
If $u=0$ then $|V_\sigma(u)| = \sigma$, and it follows from (\ref{eqn:response}) 
that $C(u) \ne R^\star_\sigma$ if $\sigma > \rho$. It has already been established
that $|R_\sigma(0,u)|$ has maxima of $R^\star_\sigma$ at $\pm\sigma$, which implies
that the response $C(u)$ will be \emph{bimodal} unless
\begin{equation}
\sigma \le \rho
\label{eqn:impulse-limit}
\end{equation}
as illustrated figure \ref{fig:impulse}. This condition is strictly imposed, as it
would be undesirable to have a bimodal response to a unimodal signal.
In general, $\rho$ should be made as large as possible for a given 
$\sigma$, in order to achieve as much shift-invariance as possible.
Recall, for example, that the least-squares approximations in section \ref{sec:error}
were demonstrated for $\rho=1.5\sigma$. 

\subsubsection{Step}

\begin{figure}[!ht]
\begin{center}
\includegraphics{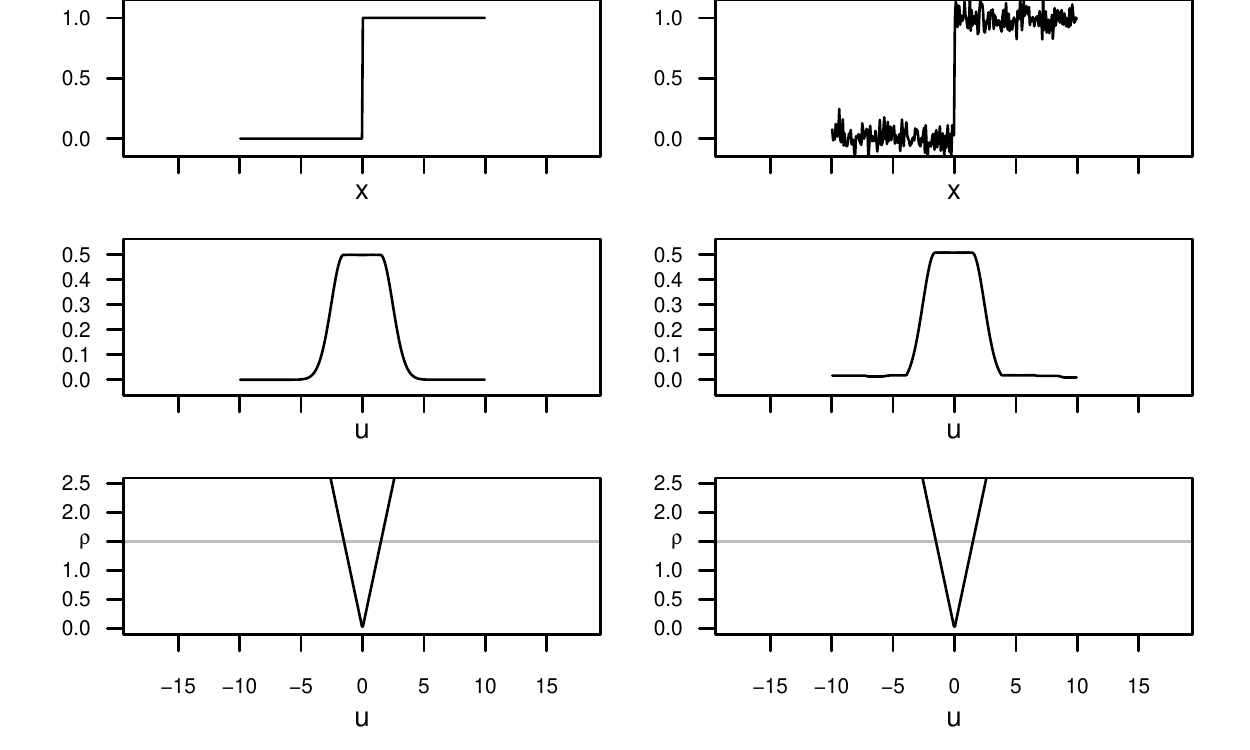}
\vspace*{-.1in}
\caption{{Step response}. \textbf{Left column:} The top plot shows the step-edge 
$S_\alpha(x)$, as defined in (\ref{eqn:step-sig}). The middle plot shows the response
$C(u)$. The bottom plot shows the distance function $V_\alpha(u)$, as in
(\ref{eqn:step-dist}). Note that the response is unconditionally unimodal in this case.
\textbf{Right column:} As before, except with Gaussian noise (SD 0.075) added 
independently at each point. The response $C(u)$ is not significantly affected.}
\label{fig:step}
\end{center}
\end{figure}

The second test signal to be considered is the unit step function. This is arguably 
the most important example, because it is the basic model for a luminance \emph{edge}. 
Indeed, the current model is optimized for the detection of step-like edges, owing
to the use of the \emph{first} derivative as the offset filter \citep{canny-1986}.
The step can be defined from the standard sign-function, as follows
\begin{equation}
S_\alpha(x) = \frac{\alpha}{2}\bigl(1+\mathrm{sgn}(x)\bigr)
\label{eqn:step-sig}
\end{equation}
The unit step function is related to the integral 
$\varPhi(u,\sigma)$ of the normalized Gaussian function $G_0^0(x,\sigma)$
in the following way: 
\begin{align}
\mathit{\Phi}(u,\sigma) 
&= \int_{-\infty}^u G^0_0(x,\sigma)\ \mathrm{d}x \\[1ex]
&= \frac{1/\alpha}{\sigma\sqrt{\pi/2}}\int_{-\infty}^\infty 
   G^1_0(x,\sigma)\mspace{3mu}S_\alpha(u-x)\ \mathrm{d}x
\end{align}
The third integral is the convolution of $G_0^1$ with $S_\alpha$, and hence 
$\mathit{\Phi}(u,\sigma)$ is proportional to the smoothed step-edge. The linear 
response is given by the derivative,
\begin{align}
R_\alpha(0,u)
&= \frac{\sigma\sqrt{\pi/2}}{1/\alpha}\,
   \frac{\mathrm{d}}{\mathrm{d}u} \, \varPhi(u,\sigma) \\[.75ex]
&= \frac{\alpha}{2} \, G(u,\sigma)
\label{eqn:step-resp}
\end{align}
This shows that the basic response is simply an un-normalized Gaussian, located at the 
step-discontinuity. The maximum response and the signed-distance
function are evidently
\begin{equation}
R^\star_\alpha = {\alpha}/{2}
\eqand V(u) = u.
\label{eqn:step-dist}
\end{equation}
Let $\bigl(u,R(u)\bigr)$ be the Cartesian coordinates of the response curve.
The final response can be constructed from
the Gaussian (\ref{eqn:step-resp}) by
inserting the plateau $(\pm\rho,\alpha/2)$
in place of the maximum point $(0,\alpha/2)$. This is illustrated in
in figure \ref{fig:step}.

\subsubsection{Cosine}

The third class of signals to be considered are the sines and cosines. These
are of central importance, owing to their role in the Fourier synthesis of more
complicated signals. Furthermore, these functions are used to construct the 
\twod\ grating patterns that are commonly used to characterize complex cells. 
It will be convenient to base the analysis on the cosine function
\begin{equation}
S_\xi(x) = \cos\bigl(2\pi \xi x\bigr)
\label{eqn:cos-sig}
\end{equation}
where $\xi$ is the frequency. 
The Fourier transforms 
$g(x) \mapsto \mathcal{F}_x[g](\eta)$
of the filter $G_0^1(x,\sigma)$ and signal $S_\xi(x)$ are
\begin{align}
\mathcal{F}_x\bigl[G_0^1\bigr](\eta) &= \sigma\sqrt{\pi/2}\, G\bigl(\eta,1/(2\pi\sigma)\bigr) \\[.75ex]
\mathcal{F}_x\bigl[S_\xi\bigr](\eta) &= \tfrac{1}{2}\bigl(\delta(\eta-\xi) + \delta(\eta+\xi)\bigr)
\end{align}
respectively, where $\eta$ is the frequency variable. The convolution $G_0^1\star S_\xi$
can be obtained from the inverse Fourier transform of the product 
$\mathcal{F}_x\bigl[G_0^1]\,\mathcal{F}_x\bigl[S_\xi\bigr]$.
The resulting cosine is attenuated by a scale-factor $\mathcal{F}_x\bigl[G_0^1\bigr](\xi)$, 
because $\mathcal{F}_x\bigl[S_\xi](\eta)$ is zero unless $|\eta|=\xi$.
Differentiating $\cos(2\pi\xi x)$ with respect to $x$ gives $-\sin(2\pi\xi x)$,
along with a second scale-factor of $2\pi\xi$. The amplitude of the linear
response is given by the product of the two scale-factors 
$2\pi\xi$ and $\mathcal{F}_x\bigl[G_0^1](\xi)$,
which can be expressed as
\[
R^\star_\xi =
\sigma\xi\pi^{3/2} \sqrt{2}\, G\bigl(\xi,1/(2\pi\sigma)\bigr).
\label{eqn:cos-const}
\]
It can be seen that the amplitude depends on the scale $\sigma$ of the filter,
as well as on the frequency $\xi$ of the signal. 
The complete linear response is given by
\[
R_\xi(0,u) = -R^\star_\xi \sin\bigl(2\pi\xi u\bigr).
\label{eqn:cos-resp}
\]
Note that a phase-shift $u_0$ can be introduced, if required, 
by substituting $u-u_0$ for $u$. 
The rectified sine $|R_\xi(0,u)|$ is another periodic function, of twice the
frequency. The peaks of this function are separated by a distance
$\frac{1}{2\xi}$, and so 
\begin{equation}
V_\xi(u) = \biggl(u\bmod \frac{1}{2\xi}\biggr) - \frac{1}{4\xi}
\label{eqn:cos-dist}
\end{equation}
is a suitable distance function for the cosine signal (\ref{eqn:cos-sig}).
The case of sine signals is analogous, with $\sin$ replaced by $\cos$ in the linear
response (\ref{eqn:cos-resp}), and $u$ replaced by $u-\frac{1}{4\xi}$ in the distance function 
(\ref{eqn:cos-dist}).

\begin{figure}[!ht]
\begin{center}
\includegraphics{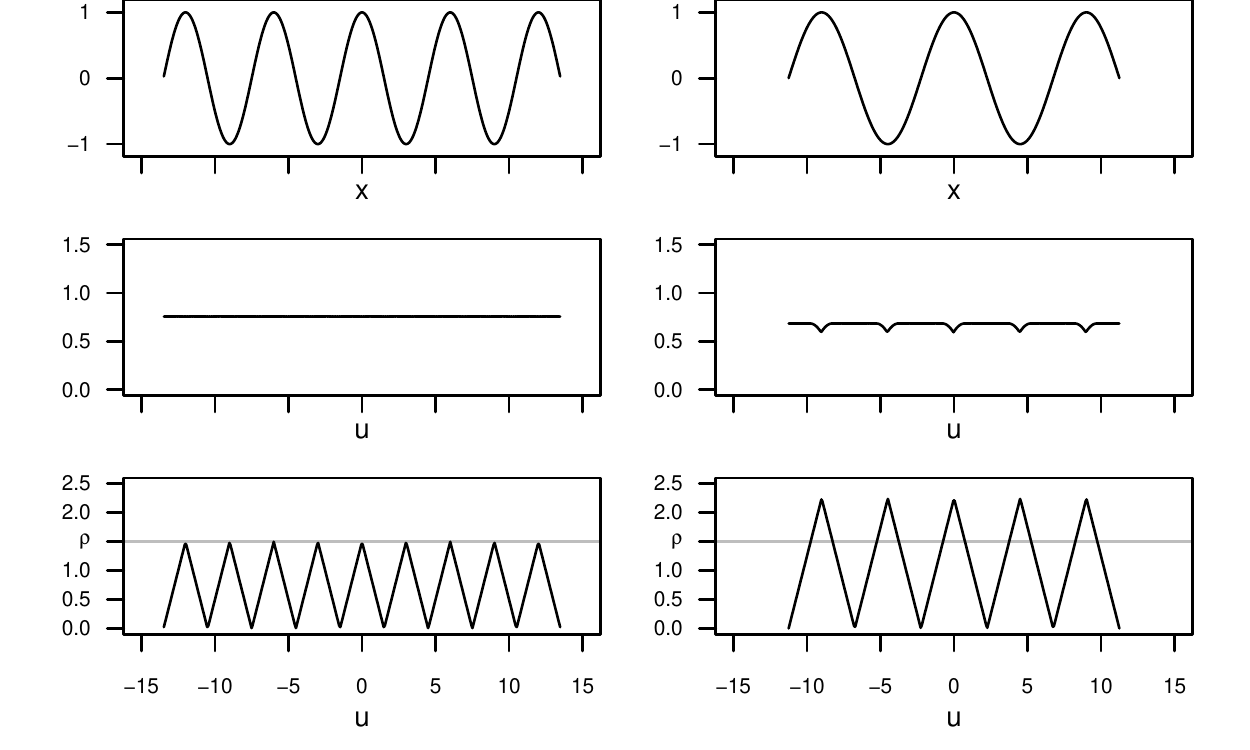}
\vspace*{-.1in}
\caption{{Cosine response}. \textbf{Left column}: The top plot shows the cosine
signal $S_\xi(x)$, as in (\ref{eqn:cos-sig}), of frequency $\xi=\frac{1}{6}$. The 
middle plot shows the response $C(u)$, which is constant. The bottom plot
shows the distance function $|V_\xi(u)|$, as in (\ref{eqn:cos-dist}). Note that the
critical case is plotted, in which the peaks of $|V_\xi(u)|$ touch the 
line $\rho=1.5\sigma$. The response is also constant for any higher frequency.
\textbf{Right column:} As before, but for a lower frequency, $\xi=\frac{1}{9}$. The distance 
function now crosses the line $\rho=1.5\sigma$, and corresponding `notches'
appear in $C(u)$.}
\label{fig:cos}
\end{center}
\end{figure}

It is important to see that the system response is \emph{entirely} constant for 
frequencies that are not too low. Specifically, the extreme values of 
(\ref{eqn:cos-dist}), with respect to $u$, are $\pm\frac{1}{4\xi}$, from which it follows 
that the response is identically $R^\star_\xi$ if $\xi \ge \frac{1}{4\rho}$. The corresponding
constraint on the wavelength $\frac{1}{\xi}$ is 
\begin{equation}
1/\xi \le 4\rho.
\label{eqn:wavelength-lim}
\end{equation}
In order to interpret this result, recall that $\rho \ge \sigma$ is required
for a unimodal impulse response (\ref{eqn:impulse-limit}). Furthermore, in section \ref{sec:error}, it
was shown that $\rho\approx 1.5\sigma$ is achievable in practice. This means that
a constant response can be expected for frequencies as low as $\xi=\frac{1}{6\sigma}$.

\subsection{Response to Natural Images}
\label{sec:images}

\begin{figure}[!ht]
\begin{center}
\includegraphics{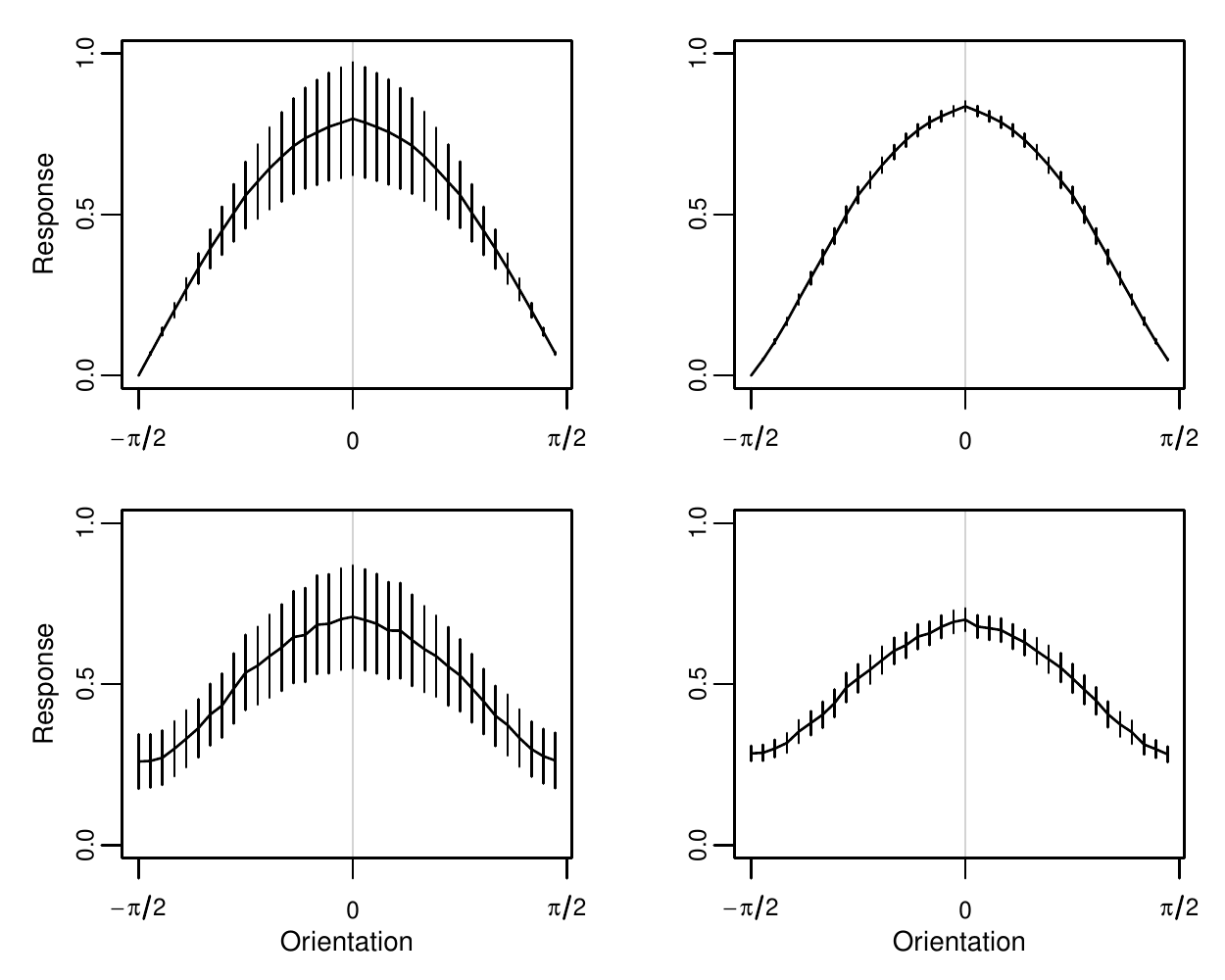}
\vspace*{-0.1in}
\caption{{Cosine response}. \textbf{Top left:} Average simple cell response $E_S(i)$
to a vertical cosine
grating of wavelength $8\sigma$. The curve indicates the mean response 
in each of 36 orientation channels $\theta_i$, and has a clear peak at zero. The
vertical bars $\pm\frac{1}{2}\sqrt{Q_S(i)}$ indicate the \textsc{rms} spatial variation 
of the response in the preferred direction of each orientation channel. 
\textbf{Top right:} Complex cell response $E_C(i)$ to the same image. The orientation tuning is 
preserved, but the variability $\pm\frac{1}{2}\sqrt{Q_C(i)}$ of the response
is greatly reduced. \textbf{Bottom left, right:} As before, but with noise added to the image. }
\label{fig:tracks-cos}
\end{center}
\end{figure}

\begin{figure}[!ht]
\begin{center}
\includegraphics{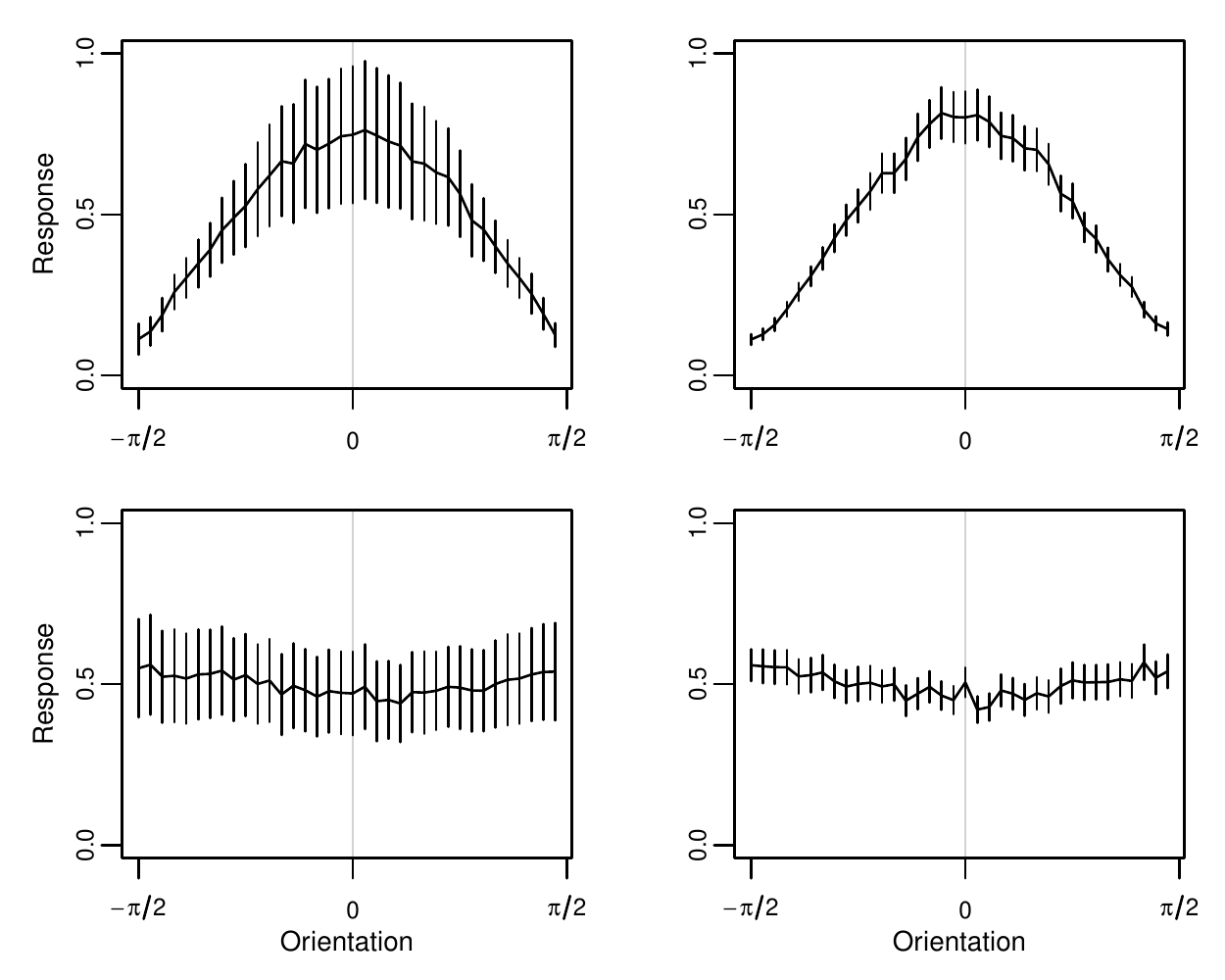}
\vspace*{-0.1in}
\caption{{Image response}. \textbf{Top:} As in figure~\ref{fig:tracks-cos}, but using
a real image that contains a dominant vertical orientation. 
Left: The simple response shows variation across all orientation channels $Q_S(i)$. 
Right: The variation of the complex response $Q_C(i)$ is much lower. \textbf{Bottom:}
As before, but using a natural image, with no dominant orientation.}
\label{fig:tracks-img}
\end{center}
\end{figure}

\noindent
This section makes a basic evaluation of the differential model, using the 
objective function of `slow feature analysis' \citep{wiskott-2002,berkes-2005}.
The procedure is as follows. Each $1024\times 768$ greyscale image is decomposed into
$i=1,\ldots,36$ orientation channels $\theta_i$ at scale $\sigma=2$~pixels. This corresponds to 
a set of simple-cell responses $S_1(\mat{x},\sigma,\theta_i)$, with an angular
separation of $5^\circ$. The steerability of $S_1$ is \emph{not} used (i.e.\ a
separate convolution is done for each $\theta_i$) in order
to minimize any angular bias in the image sampling.
A set of straight tracks 
\begin{equation}
\mat{x}_{ijk} = \mat{p}_j \pm k\Delta\times(\cos\theta_i,\, \sin\theta_i)^\tp, 
\eqwhere
j=1,\ldots,m 
\eqand 
k=0,\ldots,n
\end{equation}
is sampled from each \twod\ response. The $m=100$ random points $\mat{p}_j$ are sampled
from a uniform distribution over the image; the sign $\pm$ is also random.
The resolution $\Delta$ is set to one pixel, and the number of steps along
each path is $n=99$. This gives a total of $100^2$ samples 
from each orientation channel. The responses at non-integral positions 
$\mat{x}_{ijk}$ are obtained by bilinear interpolation.
The samples are non-negative by definition, and a global scale factor $\gamma$
is used to make the overall \mbox{$ijk$-mean} of
$\gamma S_1(\mat{x}_{ijk},\sigma,\theta_i)$ equal to $\frac{1}{2}$.
The mean simple-cell response is then computed in each orientation channel,
\begin{equation}
E_S(i) = 
\frac{1}{mn} \sum_{j=1}^{m} \sum_{k=0}^{n} \,
\gamma S_1\bigl(\mat{x}_{ijk},\sigma,\theta_i\bigr)
\end{equation}
where the scaling by $\gamma$ ensures that $\sum_i E_S(i) = \frac{1}{2}$.
The mean quadratic variation along the paths is also computed, in each orientation
channel;
\begin{equation}
Q_S(i) = \frac{1}{mn} \sum_{j=1}^{m} \sum_{k=1}^{n} \,
\Bigl|
\gamma S_1\bigl(\mat{x}_{ijk},\sigma,\theta_i\bigr) -
\gamma S_1\bigl(\mat{x}_{ij[k-1]},\sigma,\theta_i\bigr)
\Bigr|^2.
\end{equation}
The coordinates $\mat{x}_{ijk}$ and $\mat{x}_{ij[k-1]}$ represent adjacent points
(separated by $\Delta$) on the \mbox{$j$-th}
path in the \mbox{$i$-th} channel. In summary, $E_S(i)$ measures the average response
for orientation $\theta_i$, and $Q_S(i)$ measures the average spatial variation of this response
in direction $\theta_i$. Slow feature analysis finds filters that minimize the quadratic
variation.
The orientation tuning and slowness measurements are plotted in 
figures~\ref{fig:tracks-cos} and \ref{fig:tracks-img}
as a function of $\theta_i$, 
by attaching vertical bars $\pm\frac{1}{2}\sqrt{Q_S(i)}$ to each point $E_i$.
Each test is then repeated, using the complex response $C(\mat{x},\sigma,\theta)$
in place of the simple response $S_1(\mat{x},\sigma,\theta)$, giving measurements
$E_C(i)$ and $Q_C(i)$.

Three test-images with a dominant global orientation are used. Firstly, a vertical cosine 
grating, 
$I_{\cos}(x,y) = \frac{1}{2}\bigl(1+\cos(2\pi\xi x)\bigr)$. The range is set to $\rho=1.5\sigma$,
as usual, and the wavelength is set to $\frac{1}{\xi} = 8\sigma$. These values do \emph{not} satisfy the
limit (\ref{eqn:wavelength-lim}), which ensures that the complex response will not be trivial.
The simple-cell response is shown in figure~\ref{fig:tracks-cos} (top left), and two
effects should be noted. Firstly, the response is tuned to the dominant orientation $\theta=0$, 
as can be seen from the unimodal shape of the curve. Secondly, there is a large variation 
in the response when the tracks are orthogonal to the grating, as shown by the large
bars around $\theta=0$. This is because the filter falls in and out of phase with the image
as it moves horizontally.
The corresponding complex response is shown in figure~\ref{fig:tracks-cos} (top right).
It can be seen that the orientation tuning is preserved, while the response variation
is greatly reduced. Figure \ref{fig:tracks-cos} (bottom) repeats the test, but with 
noise added to the cosine grating,
$I = 0.25 \times I_{\cos} + 0.75 \times I_{\mathrm{uni}}$, where each pixel in 
$I_{\mathrm{uni}}$ is independently sampled from the uniform distribution on $[0,1]$.
This means that
a variable simple response is obtained as the filter moves in any direction, because the
image is now truly \twod. The complex 
response reduces the variation, as shown in figure~\ref{fig:tracks-cos} (bottom). 

Figure \ref{fig:tracks-img} (top) shows results for a real image
which has an orientation-structure similar to that of the grating. The results
are analogous. 
Finally, the same
test is performed on a natural image, which contains a mixture of foliage and
rocks. Figure \ref{fig:tracks-img} (bottom) shows that, although there is no 
dominant orientation in the stimulus, the complex response remains much less 
variable than the simple response.

\section{Discussion}
\label{sec:discussion}

It has been shown that a differential model of the complex cell can be constructed 
from the local jet representation. The differential model, which works naturally 
in a basis of steerable filters, can be viewed as a constrained version of the 
\cite{hubel-1962} subunit model.

\subsection{Neural Implementation}

\begin{figure}
\begin{center}
{\includegraphics{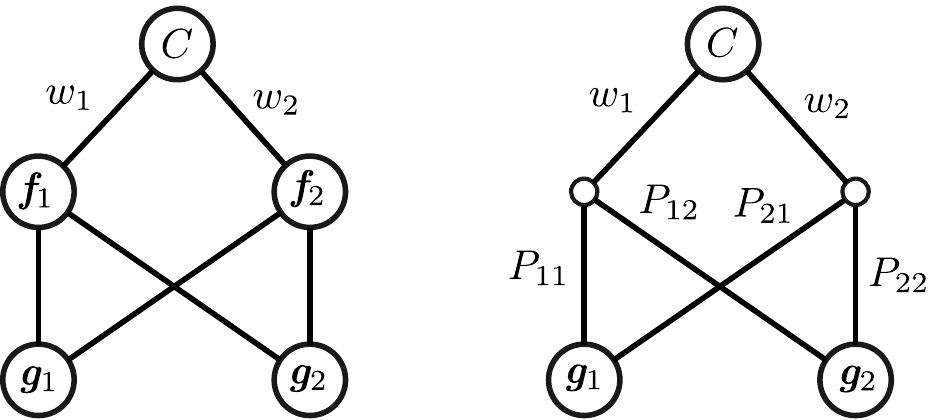}}
\caption{Two neural implementations of the complex cell $C$. These are schematic 
representations, with reduced numbers of derivative filters $\mat{g}_k$
and offset filters $\mat{f}_i$. Left: The offset filters are identified 
with an intermediate layer of simple cells. Right: The offset filters 
are implicit in the weighted sums 
$\mat{p}_1^\tp\!\mat{G}$ and $\mat{p}_2^\tp\!\mat{G}$
performed by the dendritic tree of $C$.}
\label{fig:implement}
\end{center}
\end{figure}

The qualitative components in the present approach are similar to those of the Gabor 
energy model. Both models are based on oriented linear filters, which are centred at 
the same position. Likewise, both models output a combination of the nonlinearly 
transformed filter responses.
The derivative operators $G_k(x,\sigma)$, in the differential model, are interpreted 
as simple cells, at a common location $x$ \citep{hawken-1987,young-2001a}. These are 
constructed from LGN inputs, according
to the classical model \citep{hubel-1962}. 
The offset filters $F(t_i,x)$ have two possible interpretations, 
shown in fig.~\ref{fig:implement}, as follows.

Firstly, the offset filters $F(t_i,x)$ can be interpreted as an intermediate 
layer of simple cells, each of which has an RF that is a linear combination of 
other simple cell RFs. Let the row-vector $\mat{f}_i^\tp$ represent
$F(t_i,x)$, and let $\mat{s}$ be the signal vector. It follows that the linear
response is
\begin{equation}
r_i = \mat{f}_i^\tp\!\mat{s}
\end{equation}
where $\mat{f}_i^\tp = \mat{p}_i^\tp\!\mat{G}$, according to the filter-design equation 
$\mat{F}=\mat{P}\mat{G}$ in (\ref{eqn:filt-mat}). The task of the complex cell $C$ is to 
compute the (absolute) maximum of the linear responses, $r_i$. This interpretation 
corresponds to the \textsc{hmax} model \citep{riesenhuber-1999,lampl-2004}, but 
with additional relationships imposed on the underlying simple cells.
An advantage of this interpretation is that it predicts a majority of simple cells 
with few oscillations, as is observed \citep{ringach-2002}. This follows
from the fact that the offset filters are of lower order than their derivatives, 
together with the fact that an unlimited number of offsets can be obtained 
from a given derivative basis. Furthermore, suppose that a number of complex cells 
(e.g.~of different orientations) $C_n$ are constructed from the same basis 
$\mat{G}$. A new layer of low-order offset filters $\mat{F}_n$ is required for 
each complex cell, and so the high-order filters in $\mat{G}$ are soon outnumbered
in the ensemble $\{\mat{G}, \mat{F}_1, \mat{F}_2, \ldots\}$.

An alternative physiological implementation is as follows.
Suppose that the complex cell has $i=1\ldots,M$ basal dendrites, each of which
branches out to the $K$ simple cells $G_k(x,\sigma)$. The linear response can 
then be expressed as
\begin{equation}
r_i = \mat{p}_i^\tp (\mat{G} \mat{s})
\end{equation}
where $\mat{G}\mat{s}$, the Gaussian jet response, is computed first. This interpretation requires
no intermediate simple cells. Instead, it places a fixed weight $P_{ik}$ on each
dendritic branch, and requires a summation to be performed within each dendrite.
This seems to be quite compatible with the observed cell morphology; a typical complex 
cell has a small number of basal dendrites which, unlike those of simple
cells, are extensively branched \citep{kelly-1974,gilbert-1979}.

The dendritic interpretation is more economical in the number of simple cells required, 
but less compatible with the observed simple-cell statistics \citep{ringach-2002}.
It should be noted that a mixture of the two interpretations in fig.~\ref{fig:implement}
is quite possible. For example, there could be a few intermediate simple cells,
with the remaining filters implemented by dendritic summation. In all cases, each
complex cell is associated with odd and even simple cells $\mat{g}_k$, as is observed
\citep{pollen-1981}.

The maximum (\ref{eqn:max}) over the $|r_i|$ can be approximated by a
barycentric combination, $\sum_i w_i |r_i|$.
An appropriate vector of weights can be computed as
$w_i = \mbox{$|r_i|^\beta\! \big/ (\alpha + \sum_i |r_i|^\beta)$}$.
This is a form of `softmax' \citep{bridle-1989}, with parameters $\alpha$ and $\beta$,
as described in \citep{hansard-2010}. 
Several neural models of this operation have been proposed \citep{riesenhuber-1999,yu-2002}.
For example, the $w_i$ can be interpreted as the outputs of a network of mutually
inhibitory neurons, which receive copies of the subunit responses $r_i$.
There is experimental evidence for similar arrangements, with respect to both simple
and complex cells \citep{heeger-1992,carandini-1994,lampl-2004}. 

\subsection{Experimental Predictions}
\label{sec:predictions}

This section will demonstrate that the response of the new model, to broad-band
stimuli, can easily be distinguished from that of the standard energy model.
Some qualitative predictions will also be discussed.

The response of the new model to sinusoidal stimuli is similar to that of
the energy model. Both responses are phase-invariant, provided that the
stimulus frequency is not too low (see fig.~\ref{fig:cos}). Consider, however,
a luminance step-edge that is flashed (or shifted) across the RF. The
Gabor energy is approximately Gaussian, as a function of the edge-position,
with a peak in the centre of the RF. The differential response is much 
flatter, as shown in fig.~\ref{fig:step}. This suggests that an empirical measure of 
\emph{kurtosis} could be used to distinguish between the two responses,
as will be shown below.

Let the odd-symmetric Gabor filter be defined as
$F_\parallel(x,\xi,\tau) = -G(x,\tau) \sin(2\pi\xi x)$, so that it matches the
polarity of $G_1$.
The even filter, following \citep{lehky-2005}, is defined by the numerical Hilbert
transform $F_{\!\perp} = \mathcal{H}(F_\parallel)$, in order to avoid the nonzero DC component
that arises in the cosine-based definition. The Gabor energy of a signal $S$
is then $R_F^2 = (F_\parallel\star S)^2 + (F_{\!\perp}\star S)^2$.
The envelope width $\tau$ of each Gabor filter is determined from
the constraint that the bandwidth be equal to 1.5, which is realistic for complex
cells \citep{daugman-1985}. A self-similar family of Gabor pairs, parameterized by
frequency $\xi$, can now be defined.
Quasi-Newton optimization is used to determine a frequency $\xi_0$, for which
$F_\parallel(x,\xi_0,\tau)\approx G_1(x,1)$ in the $L^2$ sense. 
Ten Gabor pairs with frequencies $\xi_k = \xi_0 2^{-k\Delta\xi}$ are constructed,
where $k=0,\ldots 9$. The corresponding differential model, with target filter
$G_1(x,\xi_0/\xi_k)$, is also constructed for each pair. Four possible ranges are 
considered for the differential models, by setting 
$\rho/\sigma = 1.0$, 1.25, 1.5, 1.75.

\begin{figure}
\begin{center}
\begin{minipage}[c]{.44\linewidth}
\centering
\vspace*{-6mm}
{\includegraphics{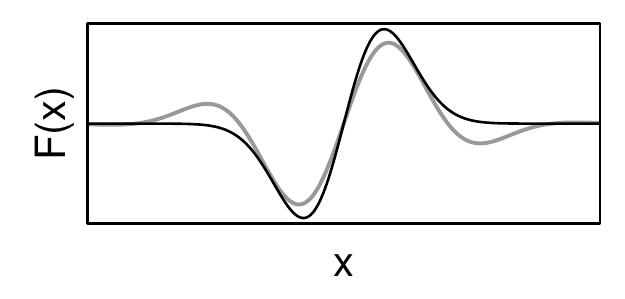}}
{\includegraphics{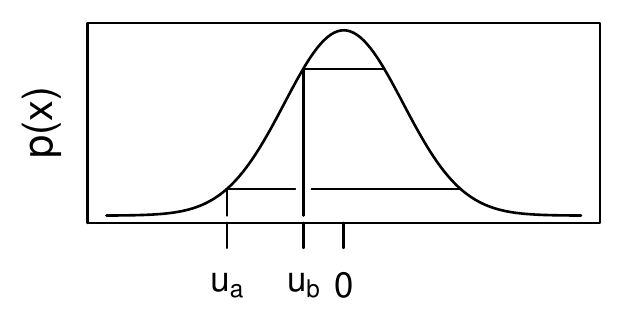}}
\end{minipage}
\hfil\begin{minipage}[c]{.52\linewidth}
\centering
{\includegraphics[scale=1]{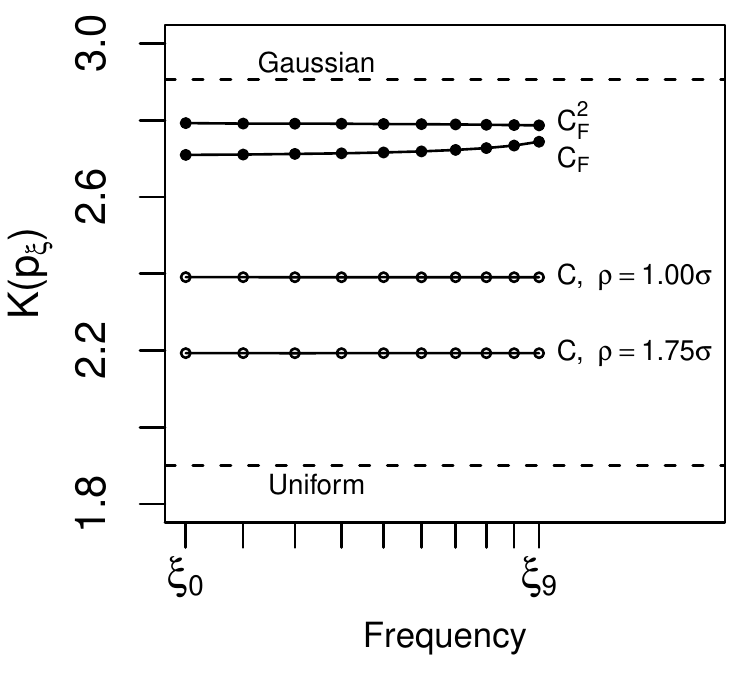}}
\end{minipage}\hfil
\vspace*{-0.1in}
\caption{Top Left: Gaussian Derivative (black) and Gabor (grey)
filters, matched subject to the bandwidth condition.
Bottom Left: The kurtosis statistic \eqref{eqn:kurtosis} is the ratio of
the two horizontal lines, shown here on a Gaussian distribution.
Right: Kurtosis of the edge-response. Each line represents a complex cell
model, parameterized by preferred frequency $\xi$. The top pair is obtained
from the Gabor energy and its square-root. The bottom pair
delimits the range of possible differential models.
The Gabor and differential responses are easily distinguished. 
The dashed lines are estimates for reference distributions.}
\label{fig:kurtosis}
\end{center}
\end{figure}

Let $p_\xi(u)$ be the response distribution, which gives the firing-rate 
for an image-edge, as a function of its offset $u$ from the centre 
$u=0$ of the complex RF. If $P_\xi(u)$ is the cumulative distribution 
$\int p_\xi(u)\,\mathrm{d}u$, then the (uncentred) kurtosis $\kappa$ of any 
$p_\xi(u)$ can be estimated \citep{crow-1967} by
\begin{equation}
K(p_\xi) = 
\frac
{P_\xi^{-1}(1-a) - P_\xi^{-1}(a)}
{P_\xi^{-1}(1-b) - P_\xi^{-1}(b)}.
\label{eqn:kurtosis}
\end{equation}
The values $a=0.025$ and $b=0.25$ are used here (making $K$ the ratio
of the 95\% and 50\% confidence intervals). 
This estimate, which can be computed by linear-interpolation
between the samples of $P_\xi(u)$, has the following interpretation.
Suppose that the offsets $u_a$ and $u_b$ are associated with
firing-rates $a$ and $b$ respectively.
The response distributions are symmetric, and so the statistic \eqref{eqn:kurtosis}
is simply the distance ratio $K(p_\xi)=u_a/u_b$, as illustrated in fig.~\ref{fig:kurtosis}.
The kurtosis could also be estimated from the empirical moments of $p_\xi(u)$,
but (\ref{eqn:kurtosis}) is much less sensitive to noise in the tails of the
distribution.

The distributions considered here lie in and around the range from the 
uniform distribution ($\kappa=\frac{9}{5}$, $K = 1.9$), to the
normal distribution ($\kappa=3$, $K\approx2.91$). It can be seen from fig.~\ref{fig:kurtosis}
that the Gabor response is approximately Gaussian, whereas the differential
responses are much flatter. Furthermore, note that the line $\rho=\sigma$ in 
fig.~\ref{fig:kurtosis} shows the \emph{maximum} kurtosis of the differential
model (determined by \ref{eqn:impulse-limit}), which is still much lower than that of the
Gabor energy. Furthermore, it can be seen that the kurtosis is approximately
independent of frequency, which simplifies the comparision. It should be noted
that a much better fit between $G_1$ and $F_\parallel$ can be obtained if
the bandwidth constraint is relaxed. This however, makes the energy responses
even more kurtotic.

The differential model makes several predictions about the configuration of simple and 
complex cells. Firstly, like the energy model, it predicts that both odd and even 
filters are required by the complex cell. Unlike the energy model, it does not 
require an exact quadrature relationship (indeed, the $G_k$ basis is not orthogonal).
More generally, an important property of the differential model is its robustness to 
deviations from the ideal simple-cell RF profiles. Derivative of Gaussian basis
was used, in the present derivation, for mathematical clarity. However, all that 
is required is a basis that spans the space of desired subunit filters $F(t_i,x)$.

The differential model also predicts a relationship between the scale 
$\sigma$ of the subunits and the radius $\rho$ of the resulting complex receptive field.
This prediction, as in the case of the energy model, is probably too strict
(i.e.\ larger complex receptive fields should be possible). However, as discussed in the
following section, the complex receptive fields can be extended by allowing multiple scales
$\sigma_j$ in the basis set of the differential model. 

A qualitative prediction of the present model is that high-order
derivative filters are required, in order to approximate the target filter over
a sufficient range $\rho$. In particular, it was shown in section \ref{sec:impulse}
that, for a unimodal impulse response, $\rho \ge \sigma$ is required.
This means, in practice, that derivative filters of order five and beyond must 
be used in the approximation, as can be seen from figure \ref{fig:errors}. 
This is interesting, because very oscillatory filters have been observed in V1 
\citep{young-2001b}. These have a natural role as high-frequency processors in 
the Gabor model. Their role is less clear in the geometric approach, because
estimates of the high-order image derivatives are of limited use. The present 
work suggests that these filters could have a different role, in providing a 
basis for spatially offset filters of low order.

\subsection{Future Directions}
\label{sec:extensions}

\noindent
There are several directions in which this model could be developed. One 
straightforward extension is to allow filters of different scales (as well
as different orders) in the basis set. Preliminary experiments confirm
that this extends the range $\rho$ of translation invariance, as would be
expected. This means that the complex cell receptive field could be made larger, 
relative to those of the underlying simple cells.
Another extension would be to allow a variety of offset-filter shapes 
(with odd, even \& mixed symmetry), rather than just the first 
derivative used here. This would lead to 
better agreement with the physiological data, which indicates a variety of
receptive field shapes among the complex subunits \citep{gaska-1987,touryan-2005,sasaki-2007}.
It would also be interesting to explore the relationship of the present work 
to the normalization model \citep{heeger-1992a,rust-2005}, and to other
models of motion and spatial processing \citep{johnston-1992,georgeson-2007}.

The present work has concentrated on local shift-invariance, because this is 
a defining characteristic of complex cells. However, mechanisms that have
other geometric invariances can be constructed in the same scale space 
framework. For example, consider the effect of a geometric scaling 
$(x,y)\rightarrow (\alpha x, \alpha y)$ on the operator
$G_0^0(x,y,\sigma,\theta)$, which represents the \mbox{$k$-th} derivative of the
normalized \twod\ Gaussian, in direction $\theta$.
The scaling $\alpha$ has no effect on the shape of the RF, as can be seen from 
the equation $G_k^0(\alpha x,\alpha y, \alpha\sigma, \theta) =
G_k^0(x,y,\sigma,\theta) \big/ \alpha^{2+k}$. This leads to simple relationships
between the responses of the RF family $G_k^0(x,y,\sigma_\ell,\theta)$, where 
$\ell=1,\ldots L$ defines a range of scales \citep{koenderink-1992,lindeberg-1998}.
Future work will consider more complicated geometric invariances (e.g.~\twod\ 
affine), in connection with the larger RFs that are found in extrastriate areas.

Another direction would be to consider how the differential model could be \emph{learned} 
from natural image data, by analogy with \citep{wiskott-2002,berkes-2005,karklin-2009}.
This could be done by fixing the local jet filters (i.e.\ simple cells), and 
then optimizing the linear transformation $\mat{P}$. The transformation
could be parameterized by coefficients $\mat{C}$, given a basis $\mat{B}$ of 
smooth functions (e.g.\ the polynomials that were used here). Alternatively, 
$\mat{P}$ could be optimized directly, subject to smoothness constraints on the 
columns $P_k(t)$. The variability of the response $C(u)$ would be a suitable 
objective function for the learning process, by analogy with slow-feature 
analysis models \citep{wiskott-2002,berkes-2005}. 
The combination of the geometric and statistical approaches to image analysis is,
more generally, a very promising aim.

\section*{Acknowledgements}
The authors would like to thank the reviewers for their help with the manuscript.


\begin{thebibliography}{}

\bibitem[Adelson and Bergen, 1985]{adelson-1985}
Adelson, E.~H. and Bergen, J.~R. (1985).
\newblock {Spatiotemporal energy models for the perception of motion}.
\newblock {\em J. Opt. Soc. Am. A}, 2(2):284--299.

\bibitem[Alonso and Martinez, 1998]{alonso-1998}
Alonso, J.-M. and Martinez, L.~M. (1998).
\newblock {Functional connectivity between simple cells and complex cells in
  cat striate cortex}.
\newblock {\em Nature Neuroscience}, 1(5):395--403.

\bibitem[Atherton, 2002]{atherton-2002}
Atherton, T.~J. (2002).
\newblock {Energy and phase orientation mechanisms: A computational model}.
\newblock {\em Spatial Vision}, 15(4):415--441.

\bibitem[Ben-Shahar and Zucker, 2004]{shahar-2004}
Ben-Shahar, O. and Zucker, S.~W. (2004).
\newblock {Geometrical Computations Explain Projection Patterns of Long Range
  Horizontal Connections in Visual Cortex}.
\newblock {\em Neural Computation}, 16(3):445--476.

\bibitem[Berkes and Wiskott, 2005]{berkes-2005}
Berkes, P. and Wiskott, L. (2005).
\newblock {Slow feature analysis yields a rich repertoire of complex cell
  properties}.
\newblock {\em Journal of Vision}, 5(6):579--602.

\bibitem[Bridle, 1989]{bridle-1989}
Bridle, J.~S. (1989).
\newblock {Probabilistic interpretation of feedforward classification network
  outputs, with relationships to statistical pattern recognition}.
\newblock In Fougelman-Soulie, F. and H\'{e}rault, J., editors, {\em
  Neuro-computing: Algorithms, Architectures and Applictions}. Springer Verlag.

\bibitem[Canny, 1986]{canny-1986}
Canny, J. (1986).
\newblock {A computational approach to edge detection}.
\newblock {\em IEEE Transactions on Pattern Analysis and Machine Intelligence},
  8(6):679--698.

\bibitem[Carandini, 2006]{carandini-2006}
Carandini, M. (2006).
\newblock {What simple and complex cells compute}.
\newblock {\em J. Physiology}, 577(2):463--466.

\bibitem[Carandini et~al., 2005]{carandini-2005}
Carandini, M., Demb, J.~B., Mante, V., Tolhurst, D.~J., Dan, Y., Olshausen,
  B.~A., Gallant, J.~L., and Rust, N. (2005).
\newblock {Do we know what the early visual system does?}
\newblock {\em J. Neuroscience}, 25:10577--10597.

\bibitem[Carandini and Heeger, 1994]{carandini-1994}
Carandini, M. and Heeger, D. (1994).
\newblock {Summation and Division by Neurons in Primate Visual Cortex}.
\newblock {\em Science}, 264:1333--1336.

\bibitem[Crow and Siddiqui, 1967]{crow-1967}
Crow, E. and Siddiqui, M. (1967).
\newblock {Robust Estimation of Location}.
\newblock {\em Journal of the American Statistical Association},
  62(318):353--389.

\bibitem[Daugman, 1985]{daugman-1985}
Daugman, J.~G. (1985).
\newblock {Uncertainty relation for resolution in space, spatial frequency, and
  orientation optimized by two-dimensional visual cortical filters}.
\newblock {\em J. Optical Soc. America}, 2(7):1160--1169.

\bibitem[Dayan and Abbott, 2001]{dayan-2001}
Dayan, P. and Abbott, L.~F. (2001).
\newblock {\em Theoretical Neuroscience}.
\newblock MIT Press.

\bibitem[Dobbins et~al., 1987]{dobbins-1987}
Dobbins, A., Zucker, S.~W., and Cynader, M.~S. (1987).
\newblock {Endstopped neurons in the visual cortex as a substrate for
  calculating curvature}.
\newblock {\em Nature}, 329:438--441.

\bibitem[Emerson et~al., 1992]{emerson-1992}
Emerson, R.~C., Bergen, J.~R., and Adelson, E.~H. (1992).
\newblock {Directionally Selective Complex Cells and the Computation of Motion
  Energy in Cat Visual Cortex}.
\newblock {\em Vision Research}, 32(2):203--218.

\bibitem[Felsberg and Sommer, 2001]{felsberg-2001}
Felsberg, M. and Sommer, G. (2001).
\newblock {The Monogenic Signal}.
\newblock {\em IEEE Transactions on Signal Processing}, 49(12):3136--3144.

\bibitem[Fleet et~al., 1996]{fleet-1996}
Fleet, D.~J., Wagner, H., and Heeger, D.~J. (1996).
\newblock {Neural encoding of binocular disparity: Energy models, position
  shifts and phase shifts}.
\newblock {\em Vision Research}, 36(12):1839--1857.

\bibitem[Freeman and Adelson, 1991]{freeman-1991}
Freeman, W.~T. and Adelson, E.~H. (1991).
\newblock {The design and use of steerable filters}.
\newblock {\em IEEE Trans. PAMI}, 13(9):891--906.

\bibitem[Gaska et~al., 1987]{gaska-1987}
Gaska, J.~P., Pollen, D.~A., and Cavanagh, P. (1987).
\newblock {Diversity of complex cell responses to even- and odd-symmetric
  luminance profiles in the visual cortex of the cat}.
\newblock {\em Experimental Brain Research}, 68:249--259.

\bibitem[Georgeson et~al., 2007]{georgeson-2007}
Georgeson, M.~A., May, K.~A., Freeman, T. C.~A., and Hesse, G.~S. (2007).
\newblock {From filters to features: Scale–space analysis of edge and blur
  coding in human vision}.
\newblock {\em Journal of Vision}, 7(13):1--21.

\bibitem[Gilbert and Wiesel, 1979]{gilbert-1979}
Gilbert, C. and Wiesel, T. (1979).
\newblock {Morphology and Intracortical Projections of Fucntionally
  Characterised Neurones in the Cat Visual Cortex}.
\newblock {\em Nature}, 280(5718):120--125.

\bibitem[Hansard and Horaud, 2010]{hansard-2010}
Hansard, M. and Horaud, R. (2010).
\newblock {Complex Cells and the Representation of Local Image-Structure}.
\newblock Technical Report 7485, INRIA.

\bibitem[Harris and Stephens, 1988]{harris-1988}
Harris, C. and Stephens, M. (1988).
\newblock {A combined corner and edge detector}.
\newblock In {\em Proc. 4th Alvey Vision Conference}, pages 147--151.

\bibitem[Hawken and Parker, 1987]{hawken-1987}
Hawken, M.~J. and Parker, A.~J. (1987).
\newblock {Spatial properties of neurons in the monkey striate cortex}.
\newblock {\em Proc.\ R.~Soc.\ Lond. B}, B~231:251--288.

\bibitem[Heeger, 1992a]{heeger-1992}
Heeger, D.~J. (1992a).
\newblock {Half-Squaring in responses of cat striate cells}.
\newblock {\em Visual Neuroscience}, 9:427--443.

\bibitem[Heeger, 1992b]{heeger-1992a}
Heeger, D.~J. (1992b).
\newblock {Normalization of Cell Responses in Cat Striate Cortex}.
\newblock {\em Visual Neuroscience}, 9:181--197.

\bibitem[Hubel and Wiesel, 1962]{hubel-1962}
Hubel, D.~H. and Wiesel, T.~N. (1962).
\newblock {Receptive fields, binocular interaction and functional architecture
  in the cat's visual cortex}.
\newblock {\em J. Physiology}, 160:106--54.

\bibitem[Johnston et~al., 1992]{johnston-1992}
Johnston, A., McOwan, P.~W., and Buxton, H. (1992).
\newblock {A computational model of the analysis of some first-order and
  second-order motion patterns by simple and complex cells}.
\newblock {\em Proc. R. Soc. Lond. B}, 250:297--306.

\bibitem[Jones and Palmer, 1987]{jones-1987}
Jones, J.~P. and Palmer, L.~A. (1987).
\newblock {An evaluation of the two-dimensional Gabor filter model of simple
  receptive fields in cat striate cortex}.
\newblock {\em J. Neurophysiology}, 58(6):1233--1258.

\bibitem[Karklin and Lewicki, 2009]{karklin-2009}
Karklin, Y. and Lewicki, M.~S. (2009).
\newblock {Emergence of complex cell properties by learning to generalize in
  natural scenes}.
\newblock {\em Nature}, 457:83--86.

\bibitem[Kelly and van Essen, 1974]{kelly-1974}
Kelly, J. and van Essen, D. (1974).
\newblock {Cell Structure and Function in the Visual Cortex of the Cat}.
\newblock {\em J. Physiology}, 238:515--547.

\bibitem[Kjaer et~al., 1997]{kjaer-1997}
Kjaer, T.~W., Gawne, T.~J., Hertz, J.~A., and Richmond, B.~J. (1997).
\newblock {Insensitivity of V1 Complex Cell Responses to Small Shifts in the
  Retinal Image of Complex Patterns}.
\newblock {\em J. Neurophysiology}, 78:3187--3197.

\bibitem[Koenderink and van Doorn, 1992]{koenderink-1992}
Koenderink, J. and van Doorn, A. (1992).
\newblock {Generic Neighborhood Operators}.
\newblock {\em IEEE Transactions on Pattern Analysis and Machine Intelligence},
  14(6):597--605.

\bibitem[Koenderink, 1984]{koenderink-1984}
Koenderink, J.~J. (1984).
\newblock {The structure of images}.
\newblock {\em Biological Cybernetics}, 50:363--370.

\bibitem[Koenderink and van Doorn, 1987]{koenderink-1987}
Koenderink, J.~J. and van Doorn, A.~J. (1987).
\newblock {Representation of local geometry in the visual system}.
\newblock {\em Biological Cybernetics}, 55:367--375.

\bibitem[Koenderink and van Doorn, 1990]{koenderink-1990}
Koenderink, J.~J. and van Doorn, A.~J. (1990).
\newblock {Receptive field families}.
\newblock {\em Biological Cybernetics}, 63:291--297.

\bibitem[Lampl et~al., 2004]{lampl-2004}
Lampl, I., Ferster, D., Poggio, T., and Riesenhuber, M. (2004).
\newblock {Intracellular Measurements of Spatial Integration and the MAX
  Operation in Complex Cells of the Cat Primary Visual Cortex}.
\newblock {\em J. Neurophysiology}, 92:2704--2713.

\bibitem[Lehky et~al., 2005]{lehky-2005}
Lehky, S.~R., Sejnowski, T.~J., and Desimone, R. (2005).
\newblock {Selectivity and sparseness in the responses of striate complex
  cells}.
\newblock {\em Vision Research}, 45:57--73.

\bibitem[Lindeberg, 1998]{lindeberg-1998}
Lindeberg, T. (1998).
\newblock {Edge Detection and Ridge Detection with Automatic Scale Selection}.
\newblock {\em International Journal of Computer Vision}, 30(2):117--154.

\bibitem[Martinez and Alonso, 2003]{martinez-2003}
Martinez, L.~M. and Alonso, J.-M. (2003).
\newblock {Complex Receptive Fields in Primary Visual Cortex}.
\newblock {\em The Neuroscientist}, 9(5):317--331.

\bibitem[\mbox{De~Valois} et~al., 1982]{devalois-1982}
\mbox{De~Valois}, R.~L., Albrecht, D.~G., and Thorell, L.~G. (1982).
\newblock {Spatial frequency selectivity of cells in macaque visual cortex}.
\newblock {\em Vision Research}, 21:545--559.

\bibitem[Mechler et~al., 2002]{mechler-2002b}
Mechler, F., Reich, D.~S., and Victor, J.~D. (2002).
\newblock {Detection and Discrimination of Relative Spatial Phase by V1
  Neurons}.
\newblock {\em J. Neuroscience}, 22(14):6129--6157.

\bibitem[Mechler and Ringach, 2002]{mechler-2002a}
Mechler, F. and Ringach, D.~L. (2002).
\newblock {On the Classification of Simple and Complex Cells}.
\newblock {\em Vision Research}, 42:1017--1013.

\bibitem[Morrone and Burr, 1988]{morrone-1988}
Morrone, M. and Burr, D. (1988).
\newblock {Feature detection in human vision: A phase dependent energy model}.
\newblock {\em Proc.\ R.~Soc.\ Lond.\ B.}, 235:221--245.

\bibitem[Movshon et~al., 1978a]{movshon-1978b}
Movshon, J.~A., Thompson, I.~D., and Tolhurst, D.~J. (1978a).
\newblock {Receptive field organization of complex cells in the cat's striate
  cortex}.
\newblock {\em J. Physiology}, 283:79--99.

\bibitem[Movshon et~al., 1978b]{movshon-1978a}
Movshon, J.~A., Thompson, I.~D., and Tolhurst, D.~J. (1978b).
\newblock {Spatial summation in the receptive fields of simple cells in the
  cat's striate cortex}.
\newblock {\em J. Physiology}, 283:53--77.

\bibitem[Orban, 2008]{orban-2008}
Orban, G.~A. (2008).
\newblock Higher order visual processing in macaque extrastriate cortex.
\newblock {\em Physiological Reviews}, 88:59--89.

\bibitem[Perona, 1995]{perona-1995}
Perona, P. (1995).
\newblock {Deformable kernels for early vision}.
\newblock {\em IEEE Trans. PAMI}, 17(5):488--499.

\bibitem[Petitot, 2003]{petitot-2003}
Petitot, J. (2003).
\newblock {The Neurogeometry of Pinwheels as a Sub-Riemannian Contact
  Structure}.
\newblock {\em J. Physiol Paris}, 97:265--309.

\bibitem[Pollen and Ronner, 1983]{pollen-1983}
Pollen, A.~D. and Ronner, S.~F. (1983).
\newblock {Visual cortical neurons as localized spatial frequency filters}.
\newblock {\em IEEE Transactions on Systems, Man \& Cybernetics}, 13:907--916.

\bibitem[Pollen and Ronner, 1981]{pollen-1981}
Pollen, D.~A. and Ronner, S.~F. (1981).
\newblock {Phase relationships between adjacent simple cells in the visual
  cortex}.
\newblock {\em Science}, 212(4501):1409--1411.

\bibitem[Press et~al., 1992]{press-1992}
Press, W.~H., Teukolsky, S.~A., Vetterling, W.~T., and Flannery, B.~P. (1992).
\newblock {\em Numerical Recipes in~C}.
\newblock Cambridge University Press, 2nd edition.

\bibitem[Riesenhuber and Poggio, 1999]{riesenhuber-1999}
Riesenhuber, M. and Poggio, T. (1999).
\newblock {Hierarchical models of object recognition in cortex}.
\newblock {\em Nature Neuroscience}, 2(11):1019--1025.

\bibitem[Ringach, 2002]{ringach-2002}
Ringach, D. (2002).
\newblock {Spatial Structure and Symmetry of Simple-Cell Receptive Fields in
  Macaque Primary Visual Cortex}.
\newblock {\em J. Neurophysiology}, 88:455--463.

\bibitem[Rust et~al., 2005]{rust-2005}
Rust, N., Schwartz, O., Movshon, J., and Simoncelli, E. (2005).
\newblock {Spatiotemporal Elements of Macaque V1 Receptive Fields}.
\newblock {\em Neuron}, 46:945--956.

\bibitem[Sasaki and Ohzawa, 2007]{sasaki-2007}
Sasaki, K.~S. and Ohzawa, I. (2007).
\newblock {Internal Spatial Organization of Receptive Fields of Complex Cells
  in the Early Visual Cortex}.
\newblock {\em J. Neurophysiology}, 98:1194--1212.

\bibitem[Simoncelli et~al., 1992]{simoncelli-1992}
Simoncelli, E.~P., Freeman, W.~T., Adelson, E.~H., and Heeger, D.~J. (1992).
\newblock {Shiftable Multiscale Transforms}.
\newblock {\em IEEE Transactions on Information Theory}, 38(2):587--607.

\bibitem[Skottun et~al., 1991]{skottun-1991}
Skottun, B.~C., Valois, R. L.~D., Grosof, D.~H., Movshon, J.~A., Albrecht,
  D.~G., and Bonds, A.~B. (1991).
\newblock {Classifying Simple and Complex Cells on the basis of Response
  Modulation}.
\newblock {\em Vision Research}, 31(7/8):1079--1086.

\bibitem[Spitzer and Hochstein, 1988]{spitzer-1988}
Spitzer, H. and Hochstein, S. (1988).
\newblock {Complex cell receptive field models}.
\newblock {\em Progress in Neurobiology}, 31:285--309.

\bibitem[Touryan et~al., 2005]{touryan-2005}
Touryan, J., Felsen, G., and Dan, Y. (2005).
\newblock {Spatial Structure of Complex Cell Receptive Fields Measured with
  Natural Images}.
\newblock {\em Neuron}, 45:781--791.

\bibitem[Wiskott and Sejnowski, 2002]{wiskott-2002}
Wiskott, L. and Sejnowski, T. (2002).
\newblock {Slow feature analysis: unsupervised learning of invariances}.
\newblock {\em Neural Computation}, 14(4):715--770.

\bibitem[Wundrich et~al., 2004]{wundrich-2004}
Wundrich, I.~J., von~der Malsburg, C., and W\"{u}rtz, R.~P. (2004).
\newblock {Image Representation by Complex Cell Responses}.
\newblock {\em Neural Computation}, 16(4):2563--2575.

\bibitem[Young and Lesperance, 2001]{young-2001b}
Young, R.~A. and Lesperance, R.~M. (2001).
\newblock {The Gaussian derivative model for spatial-temporal vision:
  II.~Cortical data}.
\newblock {\em Spatial Vision}, 14(3):321--389.

\bibitem[Young et~al., 2001]{young-2001a}
Young, R.~A., Lesperance, R.~M., and Meyer, W.~W. (2001).
\newblock {The Gaussian derivative model for spatial-temporal vision:
  I.~Cortical model}.
\newblock {\em Spatial Vision}, 14(3):261--319.

\bibitem[Yu et~al., 2002]{yu-2002}
Yu, A., Giese, M., and Poggio, T. (2002).
\newblock {Biophysiologically Plausible Implementations of the Maximum
  Operation}.
\newblock {\em Neural Computation}, 14(12):2857--2881.

\end{thebibliography}

\end{document}